\documentclass[]{emulateapj}    
\usepackage{graphicx} 
    \usepackage{epstopdf} 
    \usepackage{longtable} 
    \usepackage{natbib}
    \usepackage{chngpage}

 \RequirePackage[running]{lineno}
\begin{document}
\newcommand{\pc}{pseudocontinuum}
\newcommand{\asn}{SNR}
\newcommand{\asnm}{$\rm{\asn_{meas}}$}
\newcommand{\asnt}{$\rm{\asn_{target}}$}
\newcommand{\met}{Z}
\newcommand{\jband}{J\textendash{}band}
\newcommand{\ob}{h and $\chi$ Persei}
\newcommand{\obs}{Perseus OB-1}
\newcommand{\smalla}{a}
\newcommand{\nltezfit}{$-$0.04 $\pm$ 0.08}
\newcommand{\ltezfit}{$+$0.04 $\pm$ 0.10}
\newcommand{\clustlte}{$+$0.08 $\pm$ 0.12}
\newcommand{\clustnlte}{$-$0.03 $\pm$ 0.12}
\newcommand{\clustltecon}{$+$0.06 $\pm$ 0.14}
\newcommand{\clustnltecon}{$-$0.08 $\pm$ 0.13}
\newcommand{\logg}{log$g$}
\newcommand{\csq}{$\chi^2$}
\newcommand{\csqd}{$\chi^2_{{\rm data}}$}
\newcommand{\teff}{T$_{\rm{eff}}$}
\newcommand{\lte}{LTE}
\newcommand{\nlte}{NLTE}
\newcommand{\hii}{H\,{\sc ii}\rm}

\title{Quantitative Spectroscopic J\textendash{}band study of Red Supergiants in Perseus OB-1\footnotemark[\smalla]}
\author{
J. Zachary Gazak\altaffilmark{1},
Ben Davies\altaffilmark{2}, 
Rolf Kudritzki\altaffilmark{1,3},
Maria Bergemann\altaffilmark{4},
Bertrand Plez\altaffilmark{5}}
\bibliographystyle{apj}  

\begin{abstract}

We demonstrate how the metallicities of red supergiant (RSG) stars can be measured 
from quantitative spectroscopy down to resolutions of $\approx$ 3000 in the \jband.   
We have obtained high resolution spectra on a sample of the RSG population of \ob, a 
double cluster in the solar neighborhood.  We show that careful application of the {\sc marcs} 
model atmospheres returns measurements of \met\ consistent with solar metallicity.  Using 
two grids of synthetic spectra$-$one in pure \lte\ and one with \nlte\ calculations for the most 
important diagnostic lines$-$we measure \met\ = \ltezfit\ (\lte) and \met\ =  \nltezfit\ (\nlte) for 
the sample of eleven RSGs in the cluster.  We degrade the spectral resolution of our observations 
and find that those values remain consistent down to resolutions of less than 
$\lambda/\delta\lambda$ of 3000.  Using measurements of effective temperatures we compare 
our results with stellar evolution theory and find good agreement. We construct a synthetic 
cluster spectrum and find that analyzing this composite spectrum with single-star RSG models 
returns an accurate metallicity.  We conclude that the RSGs make ideal targets in the near 
infrared for measuring the metallicities of star forming galaxies out to 7-10 Mpc and up to ten 
times farther by observing the integrated light of unresolved super star clusters.  

\end{abstract}



\footnotetext[a]{Based in part on data collected at Subaru Telescope, which is operated by the National Astronomical Observatory of Japan.}

\altaffiltext{1}{Institute for Astronomy, University of Hawai'i, 2680 Woodlawn Dr, Honolulu, HI 96822, USA}
\altaffiltext{2}{Astrophysics Research Institute, Liverpool John Moores University, 146 Brownlow Hill, Liverpool L3 5RF, UK}
\altaffiltext{3}{University Observatory Munich, Scheinerstr. 1, D-81679 Munich, Germany}
\altaffiltext{4}{Institute of Astronomy, University of Cambridge, Madingley Road, Cambridge CB3 0HA, UK}
\altaffiltext{5}{Laboratoire Univers et Particules de Montpellier, Universit\'e Montpellier 2, CNRS, F-34095 Montpellier, France}

\maketitle

\section{Introduction}
Measuring metallicities in star-forming galaxies is a ubiquitous goal across the field of extragalactic astronomy. The evolutionary state of a galaxy is imprinted in the central metallicity and radial abundance gradient of iron- and $\alpha$-group elements.  Observed trends in these measurements across ranges of galactic mass, redshift, and environment constrain the theory of galaxy formation and chemical evolution.  
Central metallicity is dictated by galactic mass, a relationship encoded by the initial properties and evolution of these objects
\citep{1979A&A....80..155L,2004ApJ...613..898T,2008A&A...488..463M}.  Radial metallicity gradients provide a wealth of information needed to describe the complex dynamics of galaxy evolution including clustering, merging, infall, galactic winds, star formation history, and initial mass function \citep{2000MNRAS.313..338P,2004cmpe.conf..171G,2008A&A...483..401C,2009A&A...505..497Y,2009MNRAS.398..591S,2004MNRAS.349.1101D,2007MNRAS.374..323D,2008MNRAS.385.2181F,2007ApJ...655L..17B,2007MNRAS.375..673K,2009MNRAS.399..574W}.  


The pursuit of these scientific goals has been undermined by the difficulty of obtaining reliable metallicities.  Investigations tend to rely on spectroscopy of the emission lines of \hii\ regions.  
These methods require empirical calibration and choosing different commonly used calibrations yields varying and sometimes conflicting results from the same set of observations.  Both the slope and absolute scaling of metallicity are susceptible to choice of calibration: the mass-metallicity gradient across all galaxies and the radial gradients within individual galaxies can change from steep to flat while the overall metallicity can shift by a factor of up to four \citep{2008ApJ...681.1183K,2008ApJ...681..269K,2009ApJ...700..309B}.  Even the more physical ``T$_e-$based method'' (which utilizes auroral lines to remove the need for ``strong line'' calibrations) is potentially subject to biases$-$especially in the metal rich regime characteristic of the disks of all massive spiral galaxies \citep{2014arXiv1401.4437B,2005A&A...434..507S,2005A&A...441..981B,2010MNRAS.401.1375E,2012MNRAS.427.1463Z}.

One technique which avoids the uncertain calibrations of the ``strong line'' \hii\ region method is the quantitative spectroscopy of supergiant stars.  Blue supergiants have become a powerful tool for measuring metallicities, gradients, and distances to galaxies in and beyond the Local Group (WLM \textendash{} \citealt{2006ApJ...648.1007B,2008ApJ...684..118U}; NGC 3109 \textendash{} \citealt{2007ApJ...659.1198E}; IC1613 \textendash{} \citealt{2007ApJ...671.2028B}; M33 \textendash{} \citealt{2009ApJ...704.1120U}; M81 \textendash{} \citealt{2012ApJ...747...15K}).  This technique, 
while extremely promising, may also be subject to systematic uncertainties and needs to
be checked by independent methods. Moreover, it requires optical spectroscopy.  However,
next generation telescopes such as the TMT and  E\textendash{}ELT will be optimized for 
observations at infrared wavelengths, using adaptive optics supported multi object 
spectrographs. Thus, we need bright abundance tracers which radiate strongly in the IR.  Such stars--including red giants, the asymptotic giant branch, and red supergiants--will have a clear advantage in the future.

\begin{figure*}
\begin{centering}
\includegraphics[width=6in]{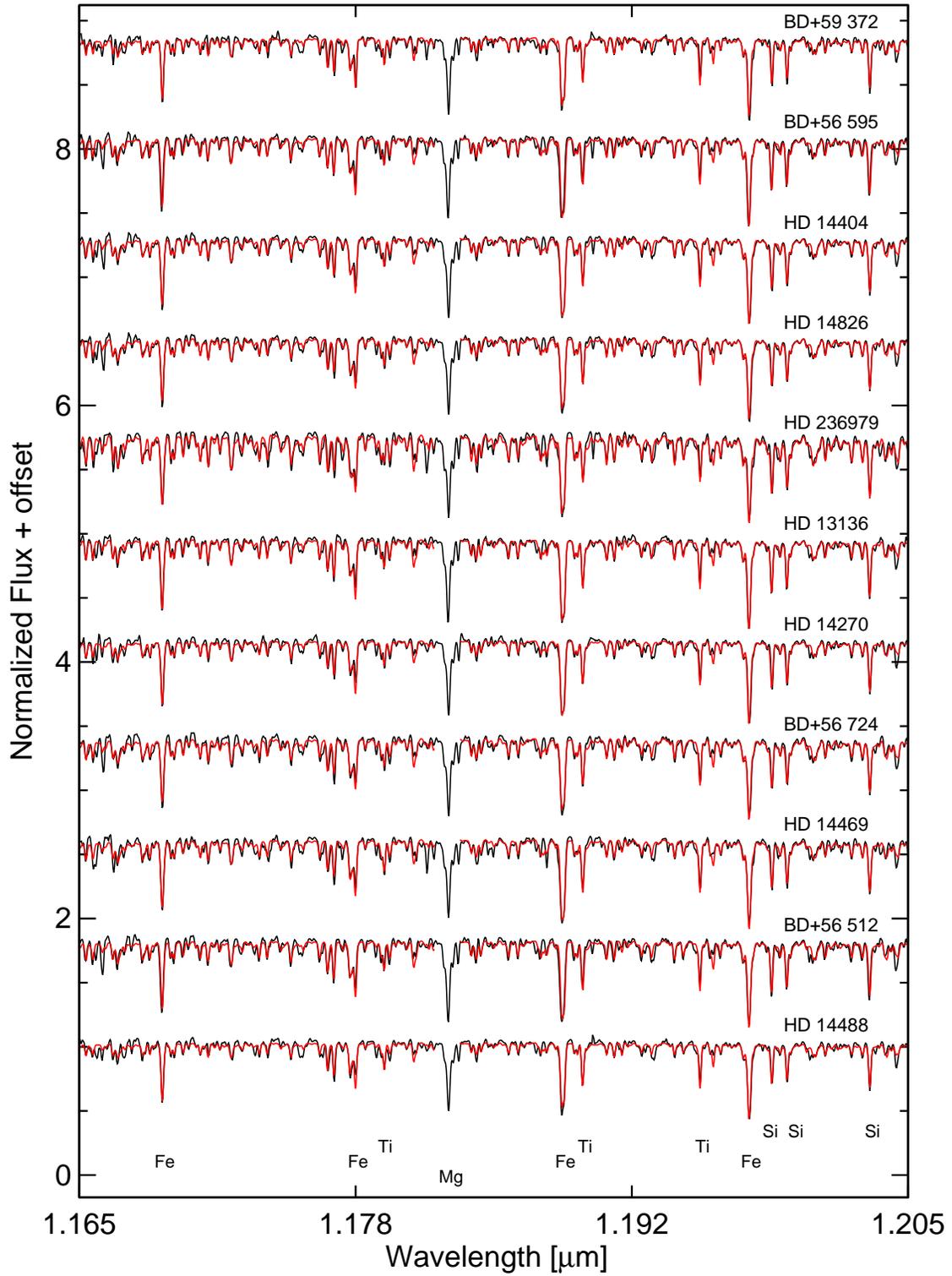}
\caption{Spectral library of RSGs observed at high resolution with IRCS on Subaru.  The main diagnostic atomic lines are labeled. Best fitting \nlte\ models are over plotted in red.  The Mg\,{\sc i} line is not included in the fit because it is calculated in \lte\ but subject to strong \nlte\ effects.  \nlte\ calculations for Mg\,{\sc i} will be implemented soon.  Plots are arranged by spectral type (see Table~\ref{tbl:perOB1}).}
\label{fig:speclib}
\end{centering}
\end{figure*}

\begin{figure*}
\begin{centering}
\includegraphics[width=6in]{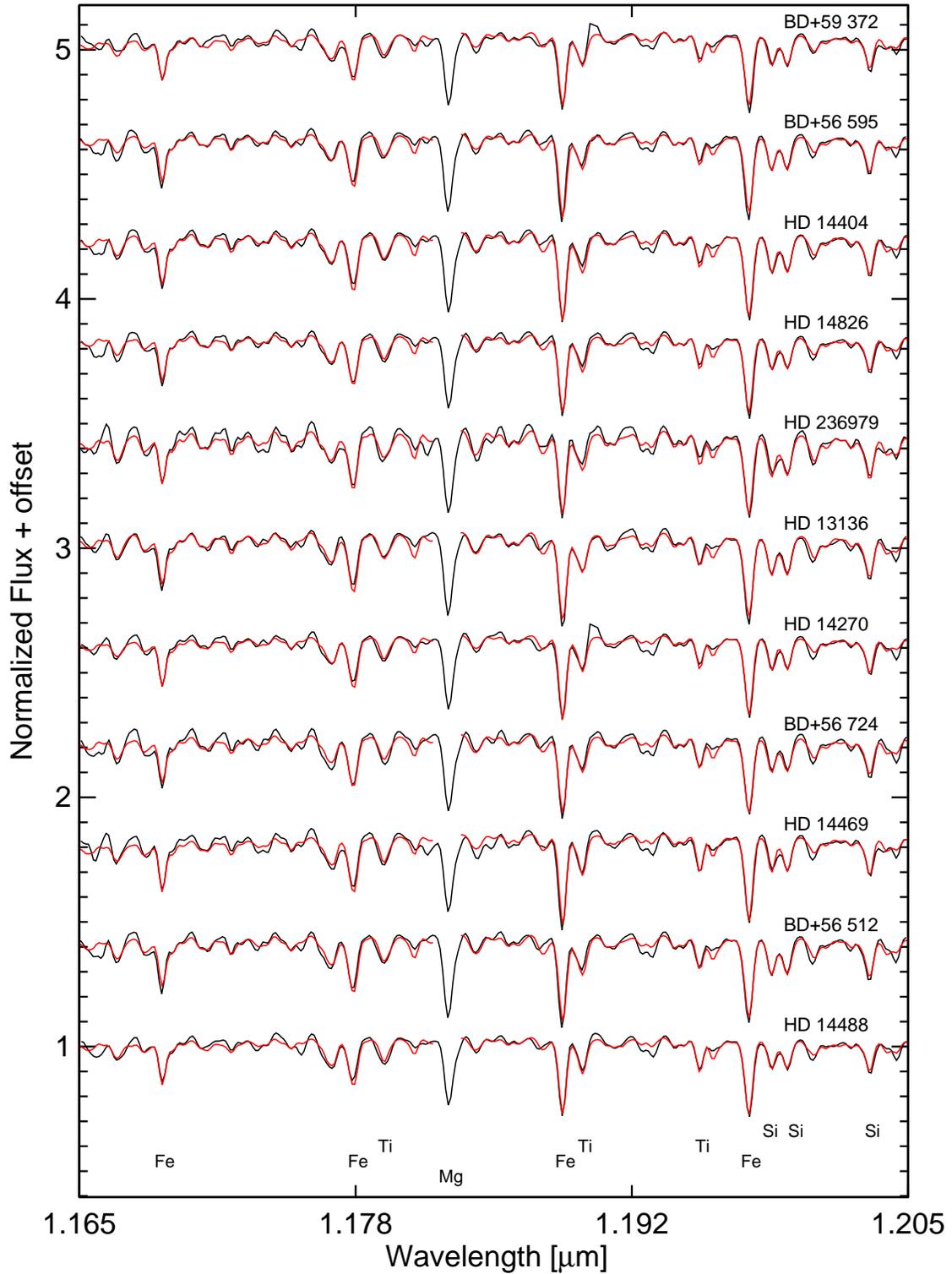}
\caption{Spectral library from Figure~\ref{fig:speclib} downgraded to a resolution of R=3000.  The main diagnostic atomic lines are labeled. Best fitting \nlte\ models are over plotted in red.  The Mg\,{\sc i} line is not included in the fit because it is calculated in \lte\ but subject to strong \nlte\ effects.  \nlte\ calculations for Mg\,{\sc i} will be implemented soon.  Plots are arranged by spectral type (see Table~\ref{tbl:perOB1}).}
\label{fig:speclib2}
\end{centering}
\end{figure*} 

\begin{figure}
\begin{centering}
\includegraphics[width=8.5cm]{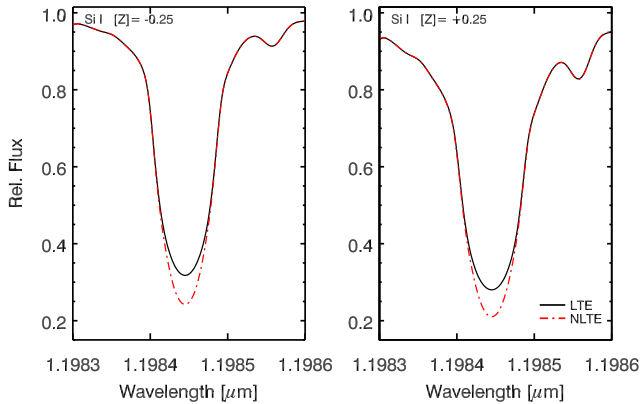}
\caption{The effects of \nlte\ corrections on diagnostic lines.  Left panel shows a model at \met\ = $-$0.25, right panel \met\ = $+$0.25}
\label{fig:lvsnl}
\end{centering}
\end{figure} 

\begin{figure}
\begin{centering}
\includegraphics[width=8.5cm]{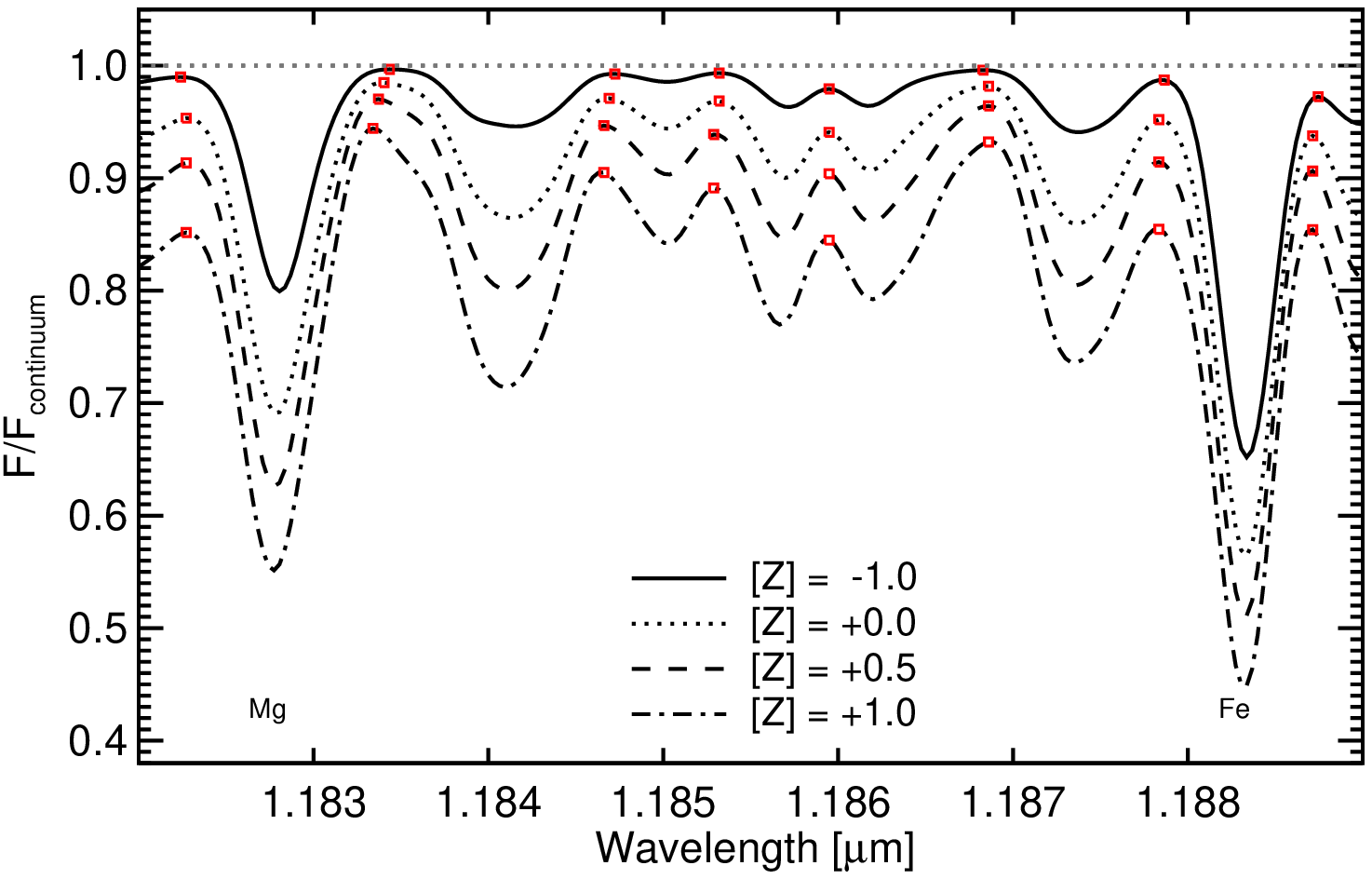} 
\includegraphics[width=8.5cm]{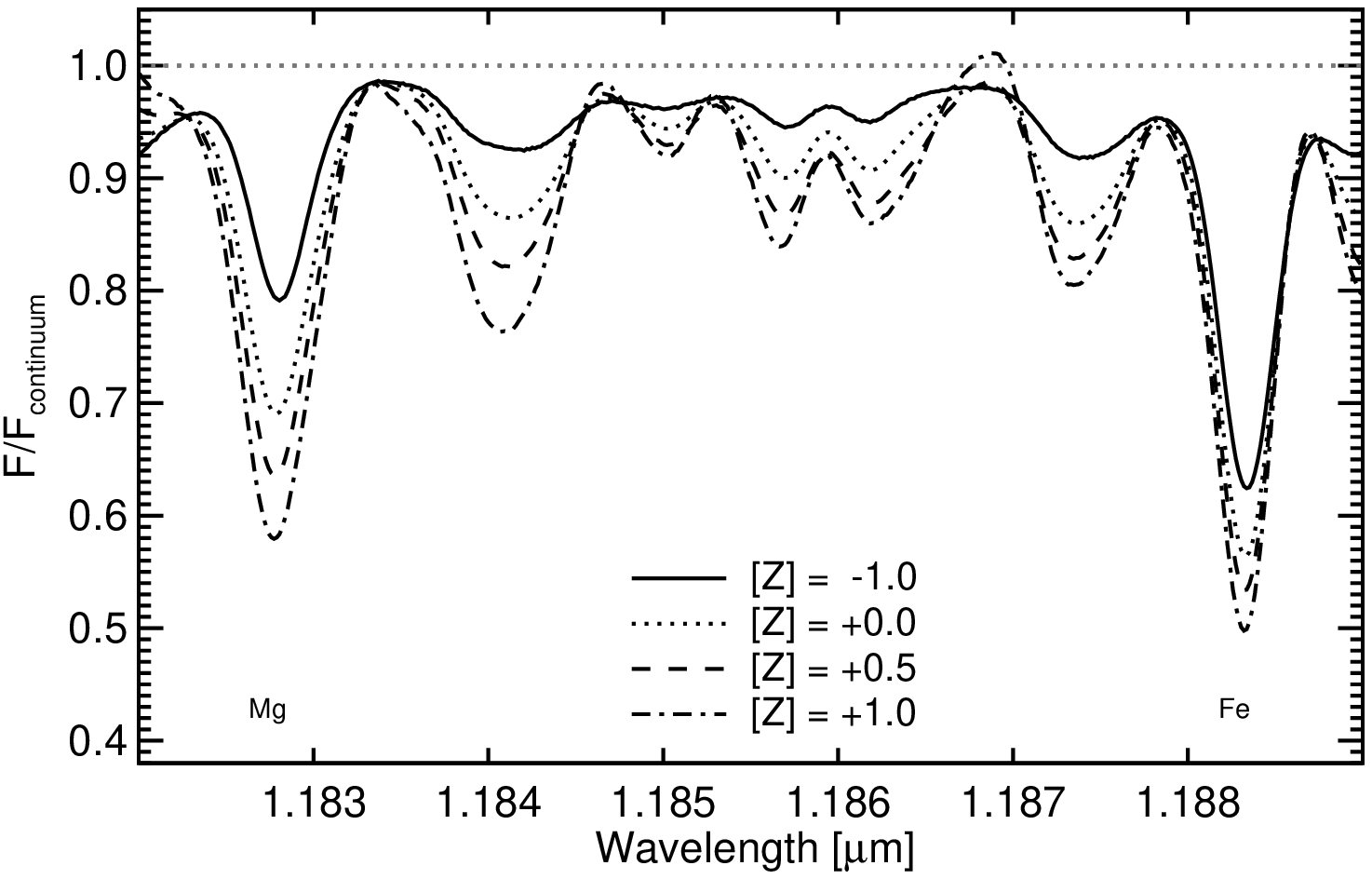}
\caption{Plot of continuum region between two strong atomic features at spectral resolution of 3000.  Each model has \teff=4000 \logg=0.0, and $\xi$ = 4.0.  Top panel: models at four values of \met\ to demonstrate the \pc.  Red squares mark the continuum points used for each model.  Bottom panel: each model is scaled to that of \met=0.0 assuming that it resembles the data set.  The variable depth of atomic spectral features as a function of metallicity is still clearly seen.  In addition, weak line features strengthen with metallicity and provide additional information with increasing metallicity.}
\label{fig:pc}
\end{centering}
\end{figure}

The extremely luminous red supergiant stars (RSGs)\textemdash{}which emit 10$^5$ to $\sim$10$^6$ L/L$_\odot$ largely in the infrared \citep{Humphreys:1979p3252}\textemdash{}thus become ideal targets for measuring extragalactic cosmic abundances.  Complications due to the densely packed spectral features synonymous with the cool, extended atmospheres of RSGs are minimized in the \jband.  Here the dominant features are isolated atomic lines of iron, titanium, silicon, and magnesium.  Molecular lines of OH, H$_2$O, CN, and CO manifest weakly or not at all in this bandpass.  A new technique proposed by \cite{2010MNRAS.407.1203D} (henceforth DFK10) has demonstrated that quantitative, medium resolution spectroscopy (R  [$\lambda/\delta\lambda$]  $\sim$2000) in the \jband\ can determine metallicities accurate to $\sim$0.15 dex for a single RSG.  While a principal limitation of the quantitative spectroscopy of stars is distance, these supergiant studies using 8-meter class telescopes have the potential to be extended to $\sim$10 Mpc \citep{2011A&A...527A..50E}.  

\begin{figure*}
\begin{centering}
\includegraphics[width=6.5in]{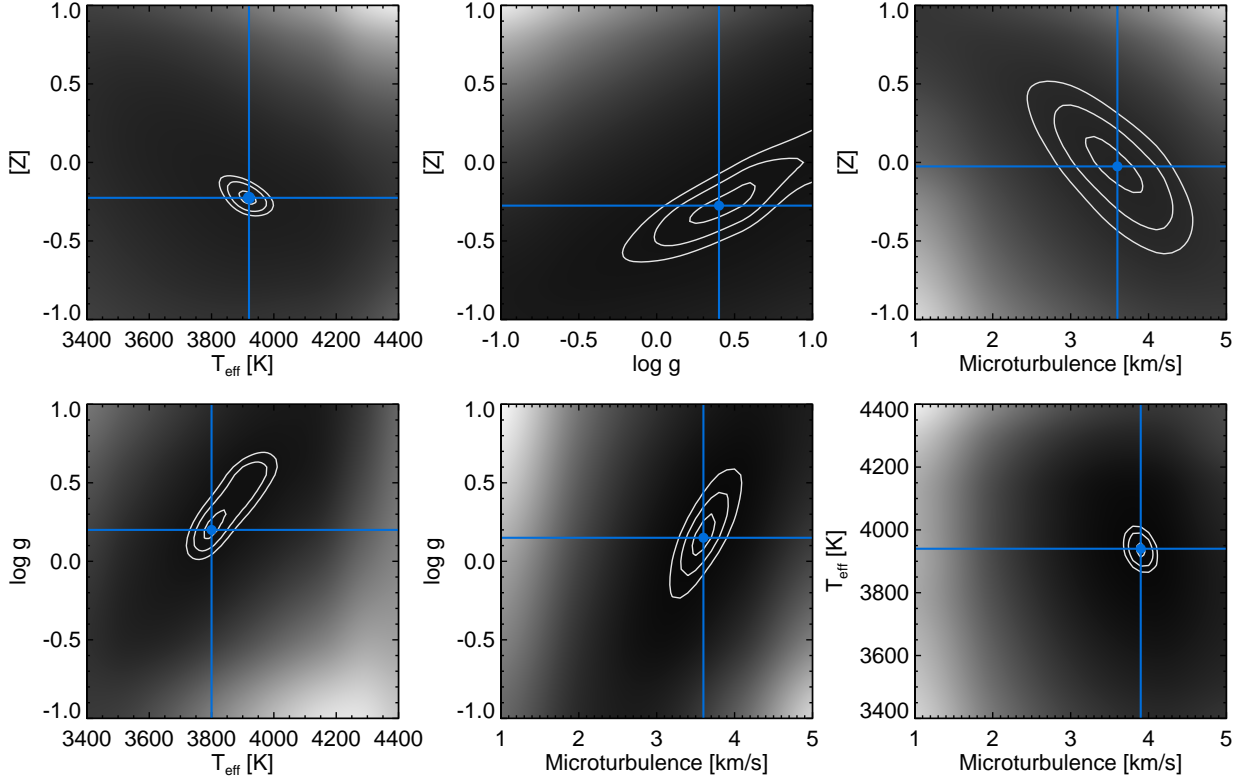}
\caption{2D contour plots used to extract fit parameters in the analysis procedure.  The smooth color gradient is an interpolated $\chi^2$ map, with dark representing lower values (better fits).  White contour lines depict fit areas of 1, 2, and 3 $\sigma$ as determined by monte carlo sampling.  The blue point at the intersection of blue lines shows the minimum $\chi^2$ in each 2D slice.}
\label{fig:fitmethod}
\end{centering}
\end{figure*} 

The \jband\ technique is thus poised to study a substantial volume of the 
local universe, one containing groups and clusters of galaxies.  The determination 
of accurate abundances for the RSG populations of star forming galaxies in this 
volume will provide an unparalleled observational constraint for models of galaxy 
formation and evolution. 
An increased utilization of supergiant stars may also aid in the proper 
development of the observationally efficient \hii-region methods while providing 
independent alternate measurement technique to the blue supergiants.

Still, DFK10 is a pilot study of the \jband\ technique and the analysis 
methods to best study these stars requires careful development and testing.  
Studies of RSGs have classically required high resolutions (R$\sim$20,000) 
in the H\textendash{}band in order to separate and study the dense 
forest of atomic and molecular features present throughout their spectra.  
Part of this requirement is driven by the scientific desire to study stellar 
evolution, for which abundances of C, N, and O are important.  The \jband\ 
technique returns no information specific to CNO processing and in exchange 
avoids the high observational overloads inherent to such studies.  This 
repurposing for extracting global chemical enrichment at modest resolution is novel.

Multiple facets of ongoing research investigate the limitations and systematic 
uncertainties of the technique in great detail.   \cite{2013ApJ...767....3D} provide 
a thorough investigation of the temperature scale of RSGs in the LMC and SMC 
and conclude that previous work at optical wavelengths measure effective temperatures
which are too cool for these RSGs.  
They find that {\sc marcs} models which fit the strong optical TiO bands produce too little
flux in the infrared to fit observed RSG spectral energy distributions.  
This discrepancy manifests in low 
measurements of effective temperature when fitting is performed with optical spectroscopy
 alone.   This problem greatly reduced in the
near-IR which correspond to deeper atmospheric layers.
Additional research is assessing the significance$-$and observational effects$-$of 
the local thermodynamic equilibrium (\lte) calculations for synthetic spectra 
produced from the {\sc marcs} models.  Departures from \lte\ have been calculated 
for iron and titanium \citep{2012ApJ...751..156B} and silicon lines 
\citep{2013ApJ...764..115B} in the \jband.  Due to the low density environments in 
the extended atmospheres of RSGs, \nlte\ effects are noticeable and can be 
significant.  For this work we have access to synthetic spectra calculated in 
both \lte\ ({\sc turbospectrum} $-$ \citealt{1998A&A...330.1109A,2012ascl.soft05004P}) 
and with iron, titanium, and silicon lines in \nlte\ using the results from 
\cite{2012ApJ...751..156B,2013ApJ...764..115B}.  

The aim of this paper is to carefully study the proposed methods of DFK10 and 
develop a proper understanding of the strengths, limitations, and systematics of 
the technique.  The ideal target for such a study is a nearby coeval population of 
RSGs in the Galaxy, such that we may study the stars as individual objects and test the potential 
of utilizing distant super star clusters (SSCs) in which the stellar population becomes
an unresolved point source.  Theoretical predictions by \cite{2013MNRAS.430L..35G} 
show that in young SSCs the RSG population dominates the near-infrared flux.  In 
this case the metallicity of the cluster could be extracted by studying the entire cluster
as a single RSG.
In order to accomplish these goals we target a galactic population of RSGs in the 
\ob\ double cluster (henceforth \obs) by performing quantitative 
spectroscopy on high resolution, high precision spectra collected using the Subaru 
Telescope atop Mauna Kea.  The presence of a large population of supergiant 
stars limits the age of \obs\ to tens of millions of years, 
and offers a laboratory for the full range of stellar astrophysics--from IMF to post-main 
sequence stellar evolution.  \cite{2010ApJS..186..191C} present a careful photometric 
and spectroscopic study of the double cluster and refine the physical parameters of 
this system.  They find an age of 14$\pm$1 Myr and estimate a minimum total stellar 
mass of 20,000 M$_\odot$.  Ages are determined using three methods which return
results in good agreement: main sequence turnoff fitting, the luminosities of red supergiants
in the clusters, and pre main sequence isochrone fitting.  
Solar metallicity is a sensible assumption for such a young 
population in the Milky Way, and studies of the B and A population of supergiant and 
giant stars$-$while incomplete$-$find solar or slightly sub-solar abundances.  Our 
high resolution spectra of eleven RSGs in \obs\ provide an ideal dataset for 
testing multiple aspects of this project.  

This paper is organized as follows.  In \S\ref{sec:obs} we discuss the observation 
and reduction of our spectral database.  \S\ref{sec:technique} contains a description 
of our atmosphere models and synthetic spectra as well as an outline of the analysis 
method we have developed.  We discuss the results of our fitting in \S\ref{sec:results} 
 We discuss and summarize the results of this work in \S\ref{sec:discussion}.

\begin{deluxetable*}{lcccccl}
\tablewidth{0pt}
\tablecaption{Perseus OB-1 Red Supergiants}
\tablehead{
\colhead{Target}
& \colhead{RA}
& \colhead{DEC}
& \colhead{m$_{\rm{V}}$}
& \colhead{m$_{\rm{J}}$}
& \colhead{m$_{\rm{H}}$}
& \colhead{SpT}}
\startdata
BD+59 372  &    01 59 39.66 & +60 15 01.9   &  9.30 & 5.33 & 4.20 &  K5-M0 I \tablenotemark{a}        \\ %
BD+56 595 &        02 23 11.03 & +57 11 58.3  &  8.18 & 4.13 & 3.22 & M1 I \tablenotemark{a}         \\ %
HD 14404   &       02 21 42.41 & +57 51 46.1   & 7.84  & 3.56 & 2.68  & M1Iab \tablenotemark{b}   \\ %
HD 14826   &  02 25 21.86 & +57 26 14.1 &  8.24 & 3.47 & 2.47   & M2 I \tablenotemark{a}      \\ 
HD 236979  &   02 38 25.42 & +57 02 46.2  &  8.10 & 3.26 & 2.30  & M2 I \tablenotemark{a}  \\
HD 13136   &   02 10 15.79 & +56 33 32.7 &  7.75 & 3.00 & 2.14   &  M2 Iab \tablenotemark{b}   \\ 
HD 14270   &     02 20 29.00 & +56 59 35.2   &  7.80 & 3.38 & 2.48  & M2.5 Iab \tablenotemark{b} \\  %
BD+56 724  & 02 50 37.89 & +56 59 00.3 &  8.70 & 3.10 & 2.00  & M3 Iab \tablenotemark{b}    \\ %
HD 14469   &     02 22 06.89 & +56 36 14.9   &  7.55 & 2.82 & 1.93  & M3-4 I \tablenotemark{a}      \\  %
BD+56 512  &  02 18 53.28 & +57 25 16.8   &  9.20 & 3.68 & 2.68  & M3 I \tablenotemark{a}      \\ %
HD 14488   &  02 22 24.30 & +57 06 34.4  &  8.50 & 3.05 & 2.11  & M4 I \tablenotemark{a}       
\enddata
\tablecomments{Target list for calibration of low resolution \jband\ RSG metallicity extraction.  m$_V$ values are adopted from \citealt{1992A&AS...94..211G}, m$_J$ and m$_H$ from 2MASS \citep{2006AJ....131.1163S}.}
\tablenotetext{a}{\,Spectral type from \cite{2005ApJ...628..973L}.}
\tablenotetext{b}{\,Spectral type from \cite{1992A&AS...94..211G}.}
\label{tbl:perOB1}
\end{deluxetable*}
 
\section{Observations}
\label{sec:obs}
On the nights of UT October 4 and 5 2011 we observed 11 of the 21 RSGs in the \obs\ 
cluster using the InfraRed Camera and Spectrograph (IRCS $-$ \citealt{ircs}) mounted 
on the Subaru telescope atop Mauna Kea.  The observations took place in non-photometric 
weather with variable partial cloud cover.  We operated to achieve maximum spectral 
resolution, using the 0$\arcsec$.14 longslit in echelle mode with natural guide star adaptive optics.  

Spectra of targets and telluric standards were bias corrected, flat fielded, extracted and 
calibrated using standard packages in IRAF.  Due to the variable cloud cover each frame 
was reduced individually and frames overwhelmed by noise were selectively removed.  
Absolute flux information cannot be recovered in such weather conditions so no flux calibrations 
were taken or used.

Observations in sub-optimal weather were possible due to the bright apparent magnitudes 
of the targets, but some spectra suffer from uncorrectable telluric contamination over certain 
wavelength ranges.  For the analysis in this paper we have masked out those spectral regions.  

A summary of the observed targets appears in Table~\ref{tbl:perOB1}, with a plot of the high resolution 
spectra in Figure~\ref{fig:speclib} and a version with the spectra artificially degraded to resolutions of
3000 in Figure~\ref{fig:speclib2}.

\section{Analysis}
\label{sec:technique}

\subsection{Atmospheric Models and Synthetic Spectra}
\label{sec:models}

For the analysis of RSGs in our sample we utilize two grids of synthetic spectra 
calculated using \lte\ and \nlte\ radiative transfer.   Both grids of model spectra 
are calculated using as input an underlying grid of {\sc marcs} model atmospheres 
\citep{2008A&A...486..951G}.  These atmospheric models are calculated in 1D 
LTE and, while not sharing the complexity of state of the art 3D models, are well 
suited for this work.  Notably, the {\sc marcs} model atmospheres have been well tested 
in the literature and converge quickly such that large grids are possible. 

The {\sc marcs} grid used in this work covers a four dimensional parameter space 
including effective temperature, log gravity, metallicity (normalized to Solar values), and microturbulence 
(\teff, log$g$, \met, $\xi$).  The dimensions of this grid can be found in Table $\ref{tbl:par_grid}$.

The grids of synthetic model spectra used in the analysis of this paper are 
calculated in first in \lte~using {\sc turbospectrum} \citep{1998A&A...330.1109A,2012ascl.soft05004P}, 
and second with \nlte~diagnostic lines (iron, titanium, and silicon) using the codes 
developed in \cite{2012ApJ...751..156B,2013ApJ...764..115B}.  See Figure~\ref{fig:lvsnl} for
a visualization of the effects of the \nlte\ corrections.

\begin{deluxetable}{lcccc}
\centering
\tabletypesize{\small}
\tablewidth{0pt}
\tablecaption{{\sc marcs} Model Grid}
\tablehead{
\colhead{Parameter}  & \colhead{Notation} & \colhead{Min}   & \colhead{Max}  & \colhead{Spacing}}
\startdata
Eff. Temperature  [K]  & \teff  & 3400  & 4000  & 100 \\
 &  &  4000  & 4400 & 200 \\
Log gravity & log$g$ & $-$0.5  & +1.0  &  0.5 \\
Metallicity [dex] & \met\  & $-$1.00 & +1.00  & 0.25 \\
Microturbulence [km/s] & $\xi$ & 1.0  & 6.0   &  1.0 
\enddata
\tablecomments{Parameter grid for {\sc marcs} atmospheres (and synthetic spectra) utilized in this work.}
\label{tbl:par_grid}
\end{deluxetable}
 
\begin{deluxetable*}{llrlrcccc}
\tablewidth{0pt}
\tabletypesize{\small}
\tablecaption{Perseus OB-1 Red Supergiants \nlte}
\tablehead{
\colhead{Target}
& \colhead{T$_{\rm{eff}}$ [K]}
& \colhead{log$g$}
& \colhead{Z [dex]}
& \colhead{$\xi$ [km/s]}
& \colhead{$\frac{\lambda}{\delta \lambda}$ \tablenotemark{a}}
& \colhead{M/M$_\odot$ } 
& \colhead{Ev. \logg}
& \colhead{Lit.\teff \tablenotemark{b}}
}
\startdata
   BD+59 372 & 3920  $\pm$   25 & +0.5 $\pm$  0.3 & $-$0.07 $\pm$  0.09 & 3.2 $\pm$ 0.2 & 13600 &   9.86 &    +0.72 & 3825 \\
   BD+56 595 & 4060 $\pm$   25 & +0.2 $\pm$  0.7 & $-$0.15 $\pm$  0.13 & 4.0 $\pm$ 0.2 & 11900 & 13.2&     +0.43  & 3800 \\
     HD 14404 & 4010  $\pm$   25 & +0.2 $\pm$  0.4 & $-$0.07 $\pm$  0.10 & 3.9 $\pm$ 0.2 & 11100 & 15.4  &   +0.24  & \nodata \\
      HD 14826 & 3930   $\pm$   26 & +0.1 $\pm$  0.2 & $-$0.08 $\pm$  0.07 & 3.7 $\pm$ 0.4 & 11200 &  15.7  &   +0.18 & 3625  \\
    HD 236979 & 4080  $\pm$   25 & $-$0.6 $\pm$  0.3 & $-$0.09 $\pm$  0.09 & 3.1 $\pm$ 0.2 & 11700 &  16.5 &    +0.18 & 3700 \\        
      HD 13136  & 4030  $\pm$   25 & +0.2 $\pm$  0.4 & $-$0.10 $\pm$  0.08 & 4.1 $\pm$ 0.2 & 12300 & 17.7  &   +0.08  & \nodata \\
      HD 14270  & 3900  $\pm$   25 & +0.3 $\pm$  0.3 & $-$0.04 $\pm$  0.09 & 3.7 $\pm$ 0.3 & 11800  & 16.2  &   +0.14 & \nodata \\
    BD+56 724  & 3840 $\pm$   25 & $-$0.4 $\pm$  0.5 & +0.08 $\pm$  0.09 & 3.0 $\pm$ 0.2 & 10900 &  16.6  &   $-$0.05 & \nodata \\
      HD 14469  & 3820  $\pm$   25 & $-$0.1 $\pm$  0.4 & $-$0.03 $\pm$  0.12 & 4.0 $\pm$ 0.2 & 10200  &17.6  &   $-$0.17  & 3575 \\
    BD+56 512  & 4090 $\pm$   35 & +0.4 $\pm$  0.3 & +0.01 $\pm$  0.12 & 4.1 $\pm$ 0.2 & 11100 &  14.7  &   +0.31 &  3600 \\
        HD 14488 & 3690 $\pm$   50 & +0.0 $\pm$  0.2 & +0.12 $\pm$  0.10 & 2.9 $\pm$ 0.2 & 10500  & 16.8 &    $-$0.07 & 3550  
\enddata
\tablenotetext{a}{\,Spectral resolution is measured to $\pm$ 100.}
\tablenotetext{b}{\,Temperatures from \cite{2005ApJ...628..973L} where target lists overlap.}

\tablecomments{Parameter fits to observed RSGs using the grid of synthetic spectra with \nlte\ corrections to Fe\,{\sc i}, Ti\,{\sc i}, and Si\,{\sc i} lines \citep{2013ApJ...764..115B,2012ApJ...751..156B}.  Masses and Evolutionary \logg\ are calculated using the Geneva stellar evolution tracks which include effects of rotation \citep{2000A&A...361..101M}. }

\label{tbl:fitsnlte}
\end{deluxetable*}

\begin{deluxetable*}{llrlrr}
\tablewidth{0pt}
\tabletypesize{\small}
\tablecaption{Perseus OB-1 Red Supergiants \lte}
\tablehead{
\colhead{Target}
& \colhead{T$_{\rm{eff}}$ [K]}
& \colhead{log$g$}
& \colhead{Z [dex]}
& \colhead{$\xi$ [km/s]}
& \colhead{$\frac{\lambda}{\delta \lambda}$}}
\startdata
           BD+59 372 & 3930 $\pm$ 90   & +0.1 $\pm$  0.3 & $-$0.10 $\pm$  0.06 & 3.4 $\pm$ 0.2 & 13400  \\
           BD+56 595 & 3970 $\pm$   25   & +0.2 $\pm$  0.3 & $-$0.08 $\pm$  0.12 & 4.1 $\pm$ 0.2 & 12400 \\
                HD 14404 & 3950 $\pm$   40  & +0.2 $\pm$  0.1 & +0.06 $\pm$  0.09 & 4.1 $\pm$ 0.2 & 11500  \\
               HD 14826 & 3870 $\pm$   25 & +0.3 $\pm$  0.2 & +0.04 $\pm$  0.10 & 3.6 $\pm$ 0.2 & 12600  \\
       HD 236979 & 4040 $\pm$  30 & $-$0.5 $\pm$  0.1 & +0.01 $\pm$  0.06 & 3.1 $\pm$ 0.2 & 12300  \\
               HD 13136 & 4030 $\pm$  40 & +0.4 $\pm$  0.2 & $-$0.11 $\pm$  0.09 & 4.3 $\pm$ 0.2 & 12200  \\
                  HD 14270 & 3890 $\pm$   25 & +0.2 $\pm$  0.3 & +0.06 $\pm$  0.12 & 3.8 $\pm$ 0.2 & 11200 \\
             BD+56 724 & 3740 $\pm$   25 & $-$0.5 $\pm$  0.1 & +0.10 $\pm$  0.06 & 3.2 $\pm$ 0.2 & 11200  \\
               HD 14469 & 3730 $\pm$   25  & $-$0.1 $\pm$  0.3 & +0.11 $\pm$  0.13 & 4.1 $\pm$ 0.2 & 10800  \\
            BD+56 512 & 3940 $\pm$   40  & +0.4 $\pm$  0.4 & +0.13 $\pm$  0.11 & 4.1 $\pm$ 0.2 & 11000 \\
     HD 14488 & 3720 $\pm$   70 & +0.2 $\pm$  0.1 & +0.17 $\pm$  0.08 & 3.2 $\pm$ 0.2 & 11700  
\enddata
\tablecomments{Parameter fits to observed RSGs using the {\sc turbospectrum} grid of synthetic spectra calculated in \lte\   \citep{2012ascl.soft05004P,1998A&A...330.1109A}. }
\label{tbl:fitslte}
\end{deluxetable*}

\subsection{Continuum Fitting}
\label{sec:contfit}

At high resolutions it is straightforward to scale a model to the continuum level 
of the data.  This is accomplished by selecting the flat regions of the spectrum, 
performing a polynomial fit to the ratio of those points to the matching observed 
flux as a function of wavelength, and then dividing the full wavelength range of 
the model by this derived fit.  In this way we are comparing the depth and shape
of spectral features between the observed spectrum and model.

At lower resolutions the effort to correct the 
continuum increases in complexity as the dense forest of weak molecular lines 
blend together to form a ``\pc'' such that the entire observation 
technically lies below continuum level.   We illustrate this effect in Figure~\ref{fig:pc} for \met = $\textendash$1.0, 0.0, $+$0.5,
and $+$1.0 at a spectral resolution of 3000.  It is not possible to know a priori how 
to then properly correct for the continuum as depth below the true continuum 
is a function of the stellar parameters themselves, especially metallicity, the primary 
target of our work. 

Our fitting method then becomes a measurement of the ratio of line depth to 
\pc\ level.  We proceed by selecting the points in any given model with normalized flux 
nearest to unity (see Figure~\ref{fig:pc}, top panel).  Assuming this is the continuum we construct an array of the ratio
of model to data fluxes at these points.   We correct for the continuum by fitting with a low order polynomial 
and applying that fit to the full model spectrum.  
In the lower panel of Figure~\ref{fig:pc} we correct each model to the \met= 0.0 model to demonstrate
that the models are not degenerate; they remain unique with respect to metallicity 
even at resolutions suffering the effects of the \pc.



\subsection{Matching Model Spectra to Data Resolution and Macroturbulence}
While a spectrograph disperses incident flux at a characteristic resolution based 
on grating and slit width, the exact spectral resolution can vary significantly 
from these expected values based on the size of large-scale turbulent motions, 
terrestrial atmospheric conditions (e.g. seeing, especially in 
the case where a point spread function is narrower than the slit width of a 
spectrograph), and instrument setup (e.g. the focus of the telescope).  For these
reasons a measurement of spectral resolution is an important component in
our analysis.  

To accomplish this, each model is degraded to a set of resolutions ranging from 
twice the expected spectral resolution and down until a $\chi^2$ minimum is passed.  
We fit a parabola through $\chi^2$ vs resolution.  The minimum of this fit is adopted 
as the best fit resolution for that model.  Upon completion we have a set of local 
minimum $\chi^2$ values each paired with a spectral resolution.  We adopt the 
model with the lowest overall $\chi^2$ as the ``best model" and the paired spectral 
resolution as the proper value for the observed data.  At this point we refit the full grid 
of models locking each at the measured best spectral resolution to calculate a 
uniform grid of $\chi^2$ values for parameter determination.

\begin{figure*}
\begin{centering}
\includegraphics[width=8.5cm]{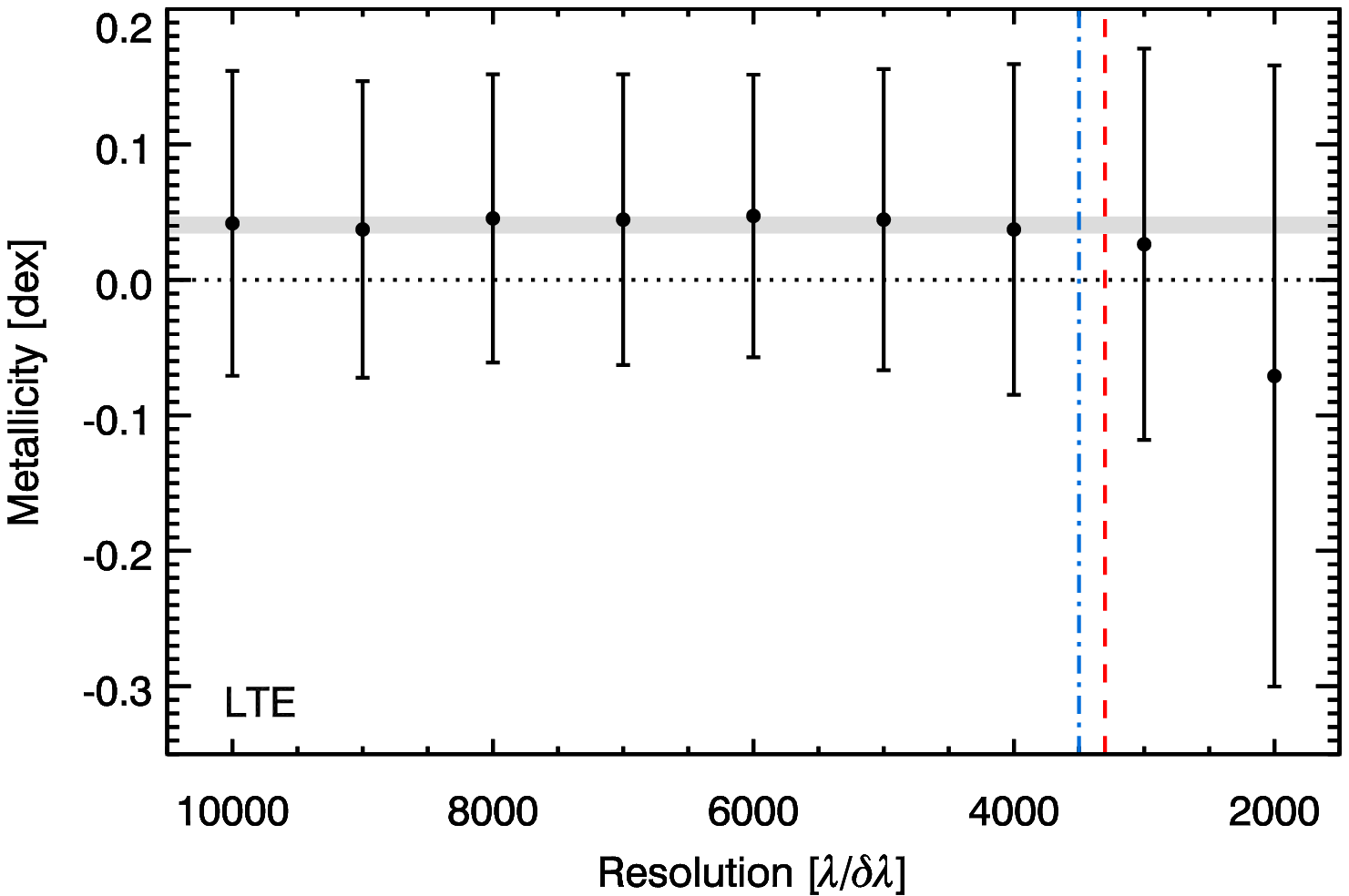} 
\includegraphics[width=8.5cm]{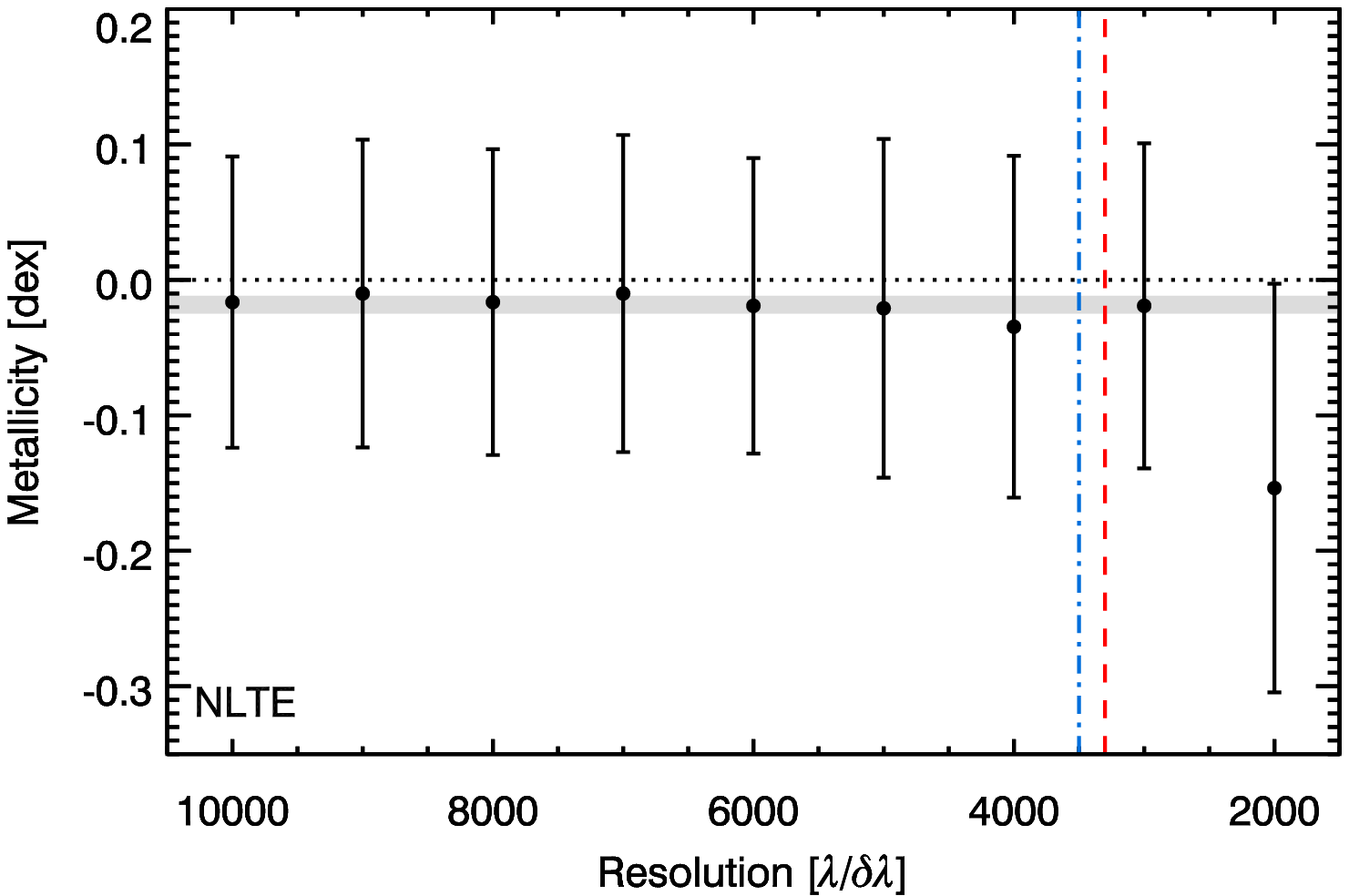}
\caption{Change in the average measured metallicity for our sample of \obs\ stars as a function of spectral resolution.  Error bars mark the standard deviation of the individual eleven measurements at each step.  The horizontal gray region shows $\pm$1$\sigma$ of the average metallicity between 12000 $\le$ R $\le $ 3000, demonstrating the stability of the technique down to resolutions of R=3000.  Vertical lines mark the spectral resolutions of key \jband\ spectrographs, KMOS on VLT in dash-dotted blue and MOSFIRE on Keck in dashed red.  A horizontal dotted line marks solar metallicity.  We plot results from the \lte\ model grid (left panel) and \nlte\ grid (right panel).  See \S\ref{sec:models} and \S\ref{sec:results2}. }
\label{fig:zvr}
\end{centering}
\end{figure*}

The expected resolution of IRCS in our particular setup is R of $\sim$20,000.  We 
measure resolutions of 11000 to 14000 (See Tables~\ref{tbl:fitsnlte} and \ref{tbl:fitslte}).  
Assuming the difference is caused by macro turbulence, we calculate expected 
$v_{\rm{macro}}$ $\approx$ 15$-$25 [km/s].  As the resolution of our observations is on 
order of 15 [km/s], we note that these values should serve only as estimates.  Still, they are 
in good agreement with literature values. 
\cite{2000ApJ...537..205R} and \cite{2007ApJ...669.1011C} find 
RSG macroturbulences of between 11-25 [km/s] using R=40,000 spectra for a population of galactic
RSGs.  

\subsection{Determination of Stellar Parameters and Errors}	
\label{sec:parerr}

After calculating a full four dimensional grid of $\chi^2$ values we extract the best fit 
parameters.  The methodology is as follows.  We begin by selecting the ``best" 
model$-$the model with the lowest $\chi^2$ value.  We use the parameters of this 
model to inform the selection of six two dimensional $\chi^2$ planes (see Figure \ref{fig:fitmethod}).
Functionally, two parameters are locked at the ``best values'' for each plane and the remaining two
parameters are varied against each other. 
We interpolate the $\chi^2$ grid of each plane onto 
a parameter grid four times as dense and take the minimum of the dense grid 
as the best fit values for the two free parameters.  After completing this procedure for
each of the six planes, we have three measured best values for each parameter.  
We average these values to arrive at a final fit for each parameter.



We assess the significance of our parameter fits with a monte carlo simulation.  
We begin by constructing a spectrum at the exact extracted parameters by linearly 
interpolating between points in the model grid.
For each of 1000 trials we add random gaussian noise of strength characteristic of the signal to noise of the measured spectrum.   We fit each noisy interpolated model as described in this section.  For each trial we determine the fit parameters and, after completing the computations, analyze the distributions of fitted values for each parameter.  In each parameter the zone of $\pm$1$\sigma$ is contained between the 15.9 and 84.1 percentile levels.  This technique accounts for the noise level in our data as well as any effects based on the spacing in our model grid.  We adopt a minimum 1$\sigma$ value of 20\%\ of the grid spacing for each parameter as we consider a fit more precise than that to be unrealistic given the possibility of nonlinear behavior between grid points.  In general our measured significance in metallicity lies above this minimum $\sigma$ value such that we may confidently trust that our grid is fine enough in metallicity space for this work.  In this work we 
find that lines of Mg\,{\sc I} are never well fit.  While the cause is under investigation, for this
analysis we mask out lines of Mg\,{\sc I} before calculating $\chi^2$.

\section{Results}
\label{sec:results} 

\subsection{The RSG population of \ob}
\label{sec:results1} 

Initial fits of the spectral database observed for this work measure a slightly sub-solar population metallicity for the \obs\ RSGs.  We measure \met\ = \ltezfit\ (\lte) and \met\ =  \nltezfit\ (\nlte), where the $\pm \sigma$ values denote the standard deviation of the sample.  Estimates of the global metallicity of the cluster are more precise, as the error in the mean scales the reported $\sigma$'s by N$^{-0.5}$ = 0.3 for our population of eleven stars.  

The \lte\ model grid measures higher metallicities for cluster stars than the \nlte\ grid.  
This is to be expected; the cores of our strongest diagnostic lines (Fe\,{\sc i}, Ti\,{\sc i}, 
and Si\,{\sc i}) are deeper in the \nlte\ case (see Figure~$\ref{fig:lvsnl}$).  For any given 
observed spectrum, a \nlte\ fit will provide a lower measurement of metallicity.  We find 
that using a fully \lte\ grid of {\sc marcs} models induces, on average, a shift in \met\ of $+0.07$
dex for RSGs near solar metallicity.  

We find good agreement between microturbulence values calculated in this work (2.9-4.3 [km/s] when compared
to high resolution spectroscopy (R$\sim$10$^5$) of $\alpha$ Ori.  \cite{2005NuPhA.758..304L} calculate a value of 4.5 
km/s using 1D ATLAS9 \lte\ models.  In \cite{2008JPhCS.130a2019W}, the authors refine that value of 3.1 km/s after fitting the same data with the newer 1D 
ATLAS12 \lte\ models.


\subsection{Parameter Stability vs. Spectral Resolution}
\label{sec:results2}

The power of the methodology presented in DFK10 is the need for only moderate resolution spectral data.  Our high resolution spectral catalogue allows for the first systematic tests of the resolution limits of the \jband\ technique.  In the following tests we degrade the resolution of our observed spectra and those spectra become the inputs to our fitting procedure.  To achieve this degradation we convolve the high resolution observed spectrum with a gaussian with characteristic width of FWHM = $\sqrt{(\lambda/\rm{R})^{2} - (\lambda/\rm{R}_{\rm{data}})^{2}}$, where R represents the output resolution of the ``new" spectra.  

We then follow identically the techniques presented in $\S\ref{sec:technique}$, treating each degraded spectra as an independent observation using nothing learned from the actual observations.  We iterate from R of 10,000 to 2,000 in steps of 1,000.  At each resolution we calculate the average and standard deviation of measured metallicity for the eleven objects.  These values have been plotted in Figure~\ref{fig:zvr}.  We find that the fitted average metallicity remains stable for both LTE and nLTE grids from spectral resolutions of 12,000 through 2,000.  Furthermore, the standard deviation in the individual metallicity measurements holds stable at $\sim$0.12 dex down to R=3000.  At this point individual atomic spectral features become too blended and diluted; the parameter fits of individual objects begin to diverge from high resolution fit values.

We conclude from these tests that the \jband\ technique can be utilized on RSGs down to spectral resolutions of 3000.  To study large populations of RSGs at extragalactic distances, then, one needs a multi object spectrograph operating at R$\ge$3000 on a telescope with enough collecting area so that the limiting magnitude is fainter than the target RSGs.  Two such instruments exist: MOSFIRE on Keck operates at R$\sim$3200 and KMOS on the VLT operates at R$\sim$3400.  These ideal instruments for the study of extragalactic populations of RSGs operate near but safely above the resolution limits of our technique.

\begin{figure}
\begin{centering}
\includegraphics[width=8.5cm]{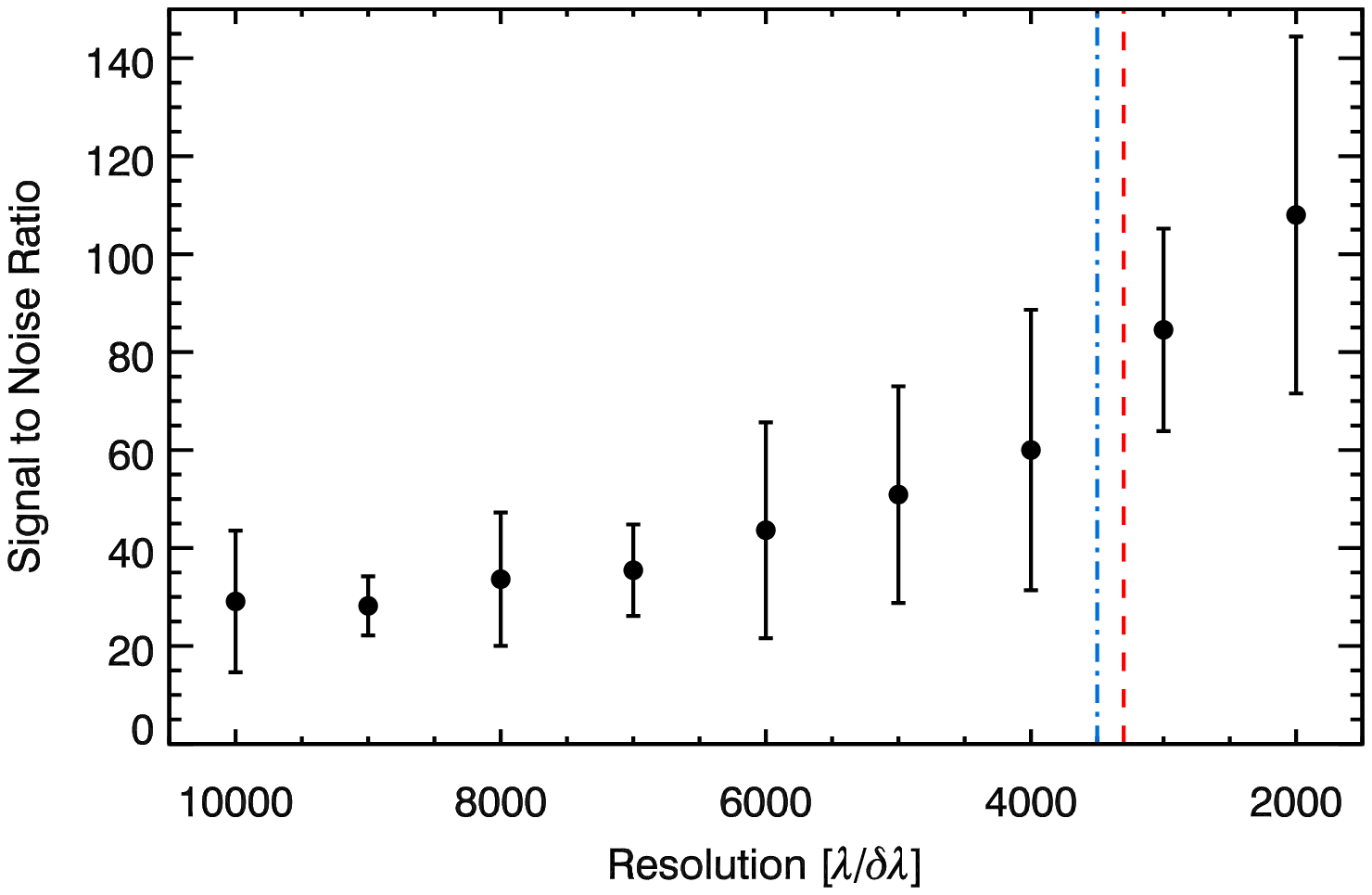} \\
\includegraphics[width=8.5cm]{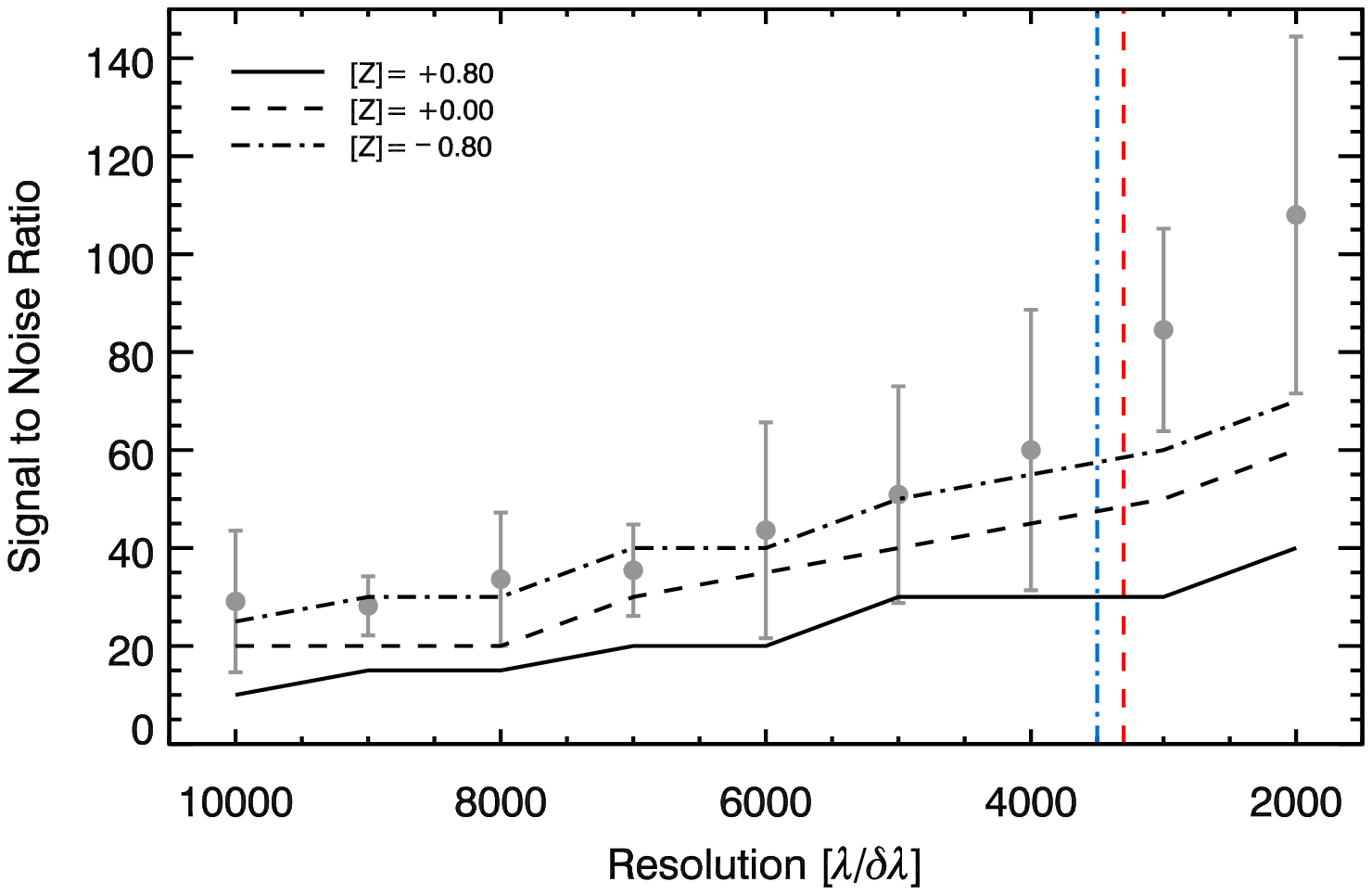}
\caption{Signal to noise ratio needed to achieve target precision for measurement of \met\ as a function of spectral resolution.  Upper Panel: Each point is the average and standard deviation of the necessary signal to noise ratio for the set of eleven RSG spectra. Lower Panel: Results for data are grayed out for comparison.  Overplotted are signal to noise predictions for models interpolated from the MARCS \nlte\ grid for metallicities of $+$0.8 (solid), 0.0 (dashed), and $-$0.8 (dash-dotted).  Vertical lines mark the spectral resolutions of key \jband\ spectrographs, KMOS on VLT in dash-dotted blue and MOSFIRE on Keck in dashed red.  For a description of the technique, see \S\ref{sec:results3}.}
\label{fig:sntest}
\end{centering}
\end{figure}

\subsection{Parameter Stability vs. Signal to Noise}
\label{sec:results3} 	

In our tests for parameter stability as a function of spectral resolution we assume that at each resolution step we have the same signal to noise ratio (\asn) as in the original, high resolution spectrum.  This value of \asn\ is $\approx$ 100-150 per object.  By reducing the spectral resolution we functionally increase the \asn\ per resolution element.  In the following, we devise a test to measure the minimum \asn\ as a function of resolution for which we obtain the same accuracy with respect to metallicity as for our high resolution case.  


We calculate the \asn\ for our original spectra, \asnm\, as follows: 

\begin{equation}
\label{eqn:scale2}
\rm{\asn_{meas}} =  \biggl[\rm{\frac{1}{N}}\sum_{i=1}^N (F_i - M_i)^2 \biggr]^{-\frac{1}{2}}
\end{equation}

In Equation~\ref{eqn:scale2}, F is the input spectrum with N points and M is the model in the grid which returns the lowest \csq.  
To measure the required \asn\ at each resolution we must adjust the effective \asn\ of the observed spectrum to any \asnt\  less than \asnm.  This is accomplished by noting that the target \asn\ is just a quadrature sum of \asnm\ and the additional noise spectrum required.  The strength of that gaussian noise, $\rm{\sigma_{scale}}$ is then 

\begin{equation}
\label{eqn:scale}
\sigma_{\rm{scale}} = \sqrt{\rm{\asn}_{\rm{target}}^{-2} - \rm{\asn}_{\rm{meas}}^{-2}}
\end{equation}




Adding random gaussian noise scaled by $\sigma_{\rm{scale}}$ to our observed spectrum degrades it to \asnt.  

To understand the \asn\ necessary to reach our target precision of $\sigma_{\rm{\met}} \approx 0.10$ we use the above method starting with a modest \asnt=5 and iteratively increase that value until the $\sigma_{\rm{\met}}$ we extract are consistent with those measured for the original spectrum, i.e., any additional \asn\ provides no increase in fit precision given the data and model grid.  The results of this test are plotted in Figure~\ref{fig:sntest}, and indicate that for instruments operating at resolutions of $\sim$3000, a \asn\ of $\sim$100 per resolution element is an ideal target for observational programs.

As with our discussion in \S\ref{sec:parerr}, we note that the residuals between data and model will not be purely gaussian in nature.  This can be due to any combination of telluric contamination, detector noise, and imperfect model atmospheres.  The \csqd\ spectrum will contain larger sporadic deviations due to those effects.  As a result, the \asnm\ of Equation~\ref{eqn:scale2} will slightly underestimate the actual \asn\ of the data.  When this propagates into Equation~$\ref{eqn:scale}$, we end up adding too little noise and not quite reaching \asnt.  This means that the curve in Figure~\ref{fig:sntest} may be skewed downwards.  The amplitude of this effect will vary due to the specifics of each observation.  This likely accounts for the scatter present in Figure~\ref{fig:sntest}.  The overall shift must be small due to the quality of our data and spectral fits.  Still, we recommend using the upper limits of the error bars in Figure~\ref{fig:sntest} as a target \asn\ when planning observations.

We perform a final experiment to test the effects of metallicity on \asn\ requirements.  We interpolate three models from our \nlte\ grid to values between grid points (\met = $+$0.8, 0.0, and $-$0.8) and reanalyze each model as described earlier in this section.  The results are plotted in the lower panel of Figure~\ref{fig:sntest}.  A trend of increasing \asn\ requirements with decreasing metallicity is indeed seen.  The effect is not overwhelming, with less than a factor of two difference between models at \met=$-$0.8 and $+$0.8.  The \asn\ required for this set of spectra are globally lower than the case of the actual data.  This is to be expected as the experiment is performed starting with perfect models which show no contamination from non-gaussian noise sources. 


\begin{figure}
\begin{centering}
\includegraphics[width=8.5cm]{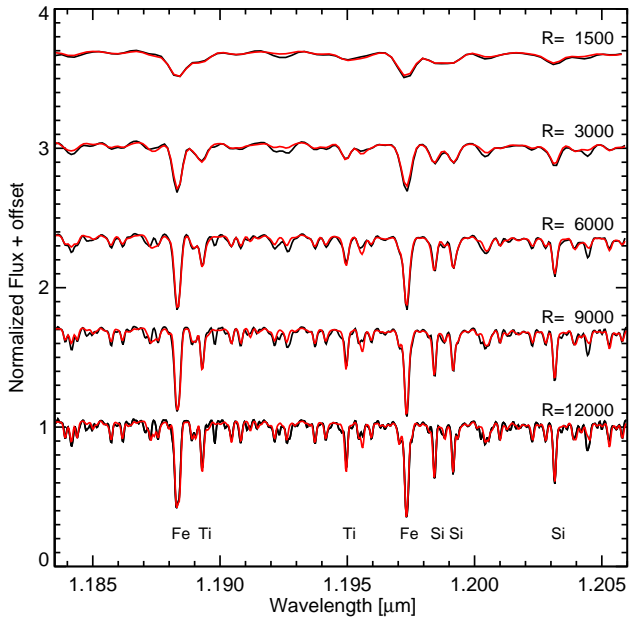}
\caption{For the analysis in this work we degrade the observed spectra by convolving those spectra with a gaussian function.  The plot shows the effects of spectral degradation on the spectrum of BD+56 595, and each is over plotted in red with the best model spectrum for that data.}
\label{fig:res}
\end{centering}
\end{figure}

\section{Discussion}
\label{sec:discussion}

\subsection{Metallicity}

The average metallicity of \met\ = \nltezfit\ obtained for
the \obs\ RSGs in this work agrees well with the 
metallicity of young massive stars in 
the solar neighborhood. \cite{2012A&A...539A.143N} studied 
a large sample early B dwarfs and giants 
using strongly improved detailed nLTE line diagnostics. 
They obtained surprisingly narrow ($\sigma \approx 0.05$)
abundance distributions for the elements C, N, O, Ne, Mg, 
Si, Fe with average values very close to the sun \citep{2009ARA&A..47..481A}. This implies that there is 
little scatter in metallicity of the young massive star
population around the sun and also practically no 
chemical evolution over the last 5 Gyrs. The fact that we
also obtain a metallicity very close to the solar value 
is, thus, a strong indication that the spectroscopic 
\jband\ method leads to reliable results.

Unfortunately, the study by \cite{2012A&A...539A.143N} does not 
include objects in \obs. However, \cite{2012A&A...543A..80F} have recently analyzed A supergiant 
stars in the solar neighborhood including some objects in
\obs. While this work does focus on the determination
of stellar parameters and does not provide a detailed 
abundance study, it provides magnesium abundances for 
three objects with an average value -0.10 dex below the 
 \cite{2012A&A...539A.143N} average of B stars in the solar 
neighborhood (the uncerainty for each individual A 
supergiant is $\approx \pm$0.07 dex). We take this as a 
confirmation that the metallicity of \obs\ is close to 
solar.

%

\begin{figure*}
\begin{centering}
\includegraphics[width=17cm]{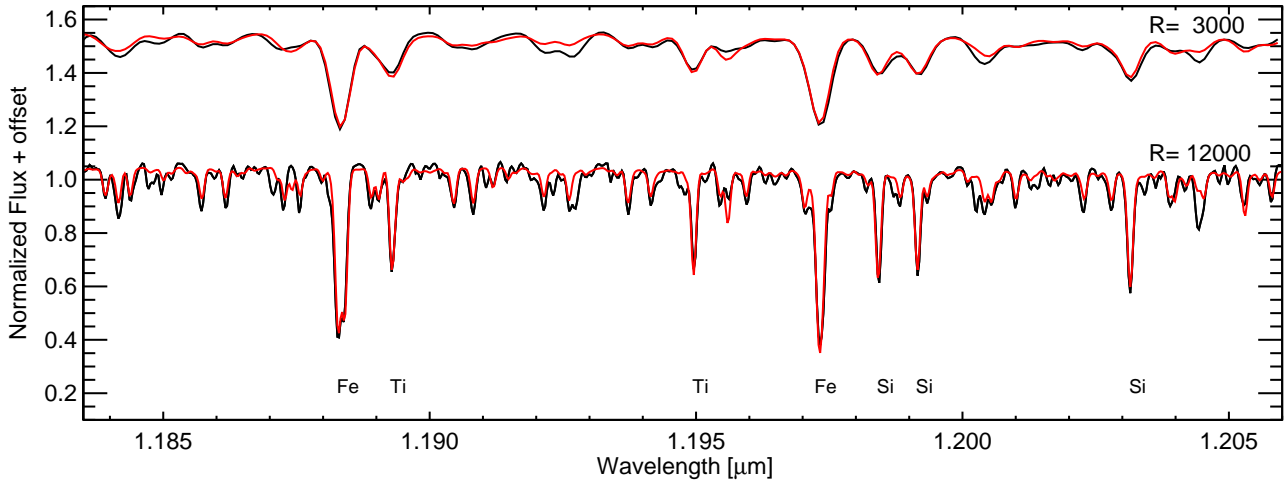}
\caption{The resulting ``cluster spectrum'' created when all eleven RSG spectra are summed together as weighted by their J magnitudes.  The spectrum is plotted twice, and on the lower spectrum we over plot the best fitting model in red.}
\label{fig:clusts}
\end{centering}
\end{figure*}

\subsection{Effective Temperatures}

We measure higher \teff\ for all stars which overlap the target list of 
\cite{2005ApJ...628..973L} (see Table~\ref{tbl:fitsnlte}).  The average 
difference in temperatures is 270 $\pm$ 130 [K] for our \nlte\ calculations (220 $\pm$ 100 when compared
with our \lte\ calculations), a significant discrepancy.
There are a number of differences between this work and that of 
\cite{2005ApJ...628..973L}, including the new \nlte\ corrections we 
use when computing synthetic spectra, the fact that we fit for \met, micro turbulence, 
and spectral resolution, and the near IR spectral window of this work.  The latter
is the most likely candidate for the large difference in measured \teff\ values.
While we work 
in the \jband, \cite{2005ApJ...628..973L} use optical spectra, concentrating on the 
strength of molecular bands of TiO to derive temperatures.  \cite{2013ApJ...767....3D} 
have shown that the derivation of RSG temperatures using quantitative spectroscopy 
in optical bandpasses returns lower values than are measured using methods which 
are less dependent on model atmospheres (the {\it flux integration method}).  In 
addition, \cite{2013ApJ...767....3D} show that
optical temperatures over predict the IR flux of RSGs when full spectral energy distributions are
available and under predict reddening as compared to nearby stars.  Temperatures derived
from near IR spectroscopy alone are closer to those values from the flux integration method.  

New work with 3D models of RSGs will likely do much to resolve the issue of temperature 
derivation for RSGs, but only a few of these models are available so far (see, for example, 
\citealt{2011A&A...535A..22C}).


\subsection{Stellar Evolution}
\label{sec:stev}
\begin{figure}
\begin{centering}
\includegraphics[width=8.5cm]{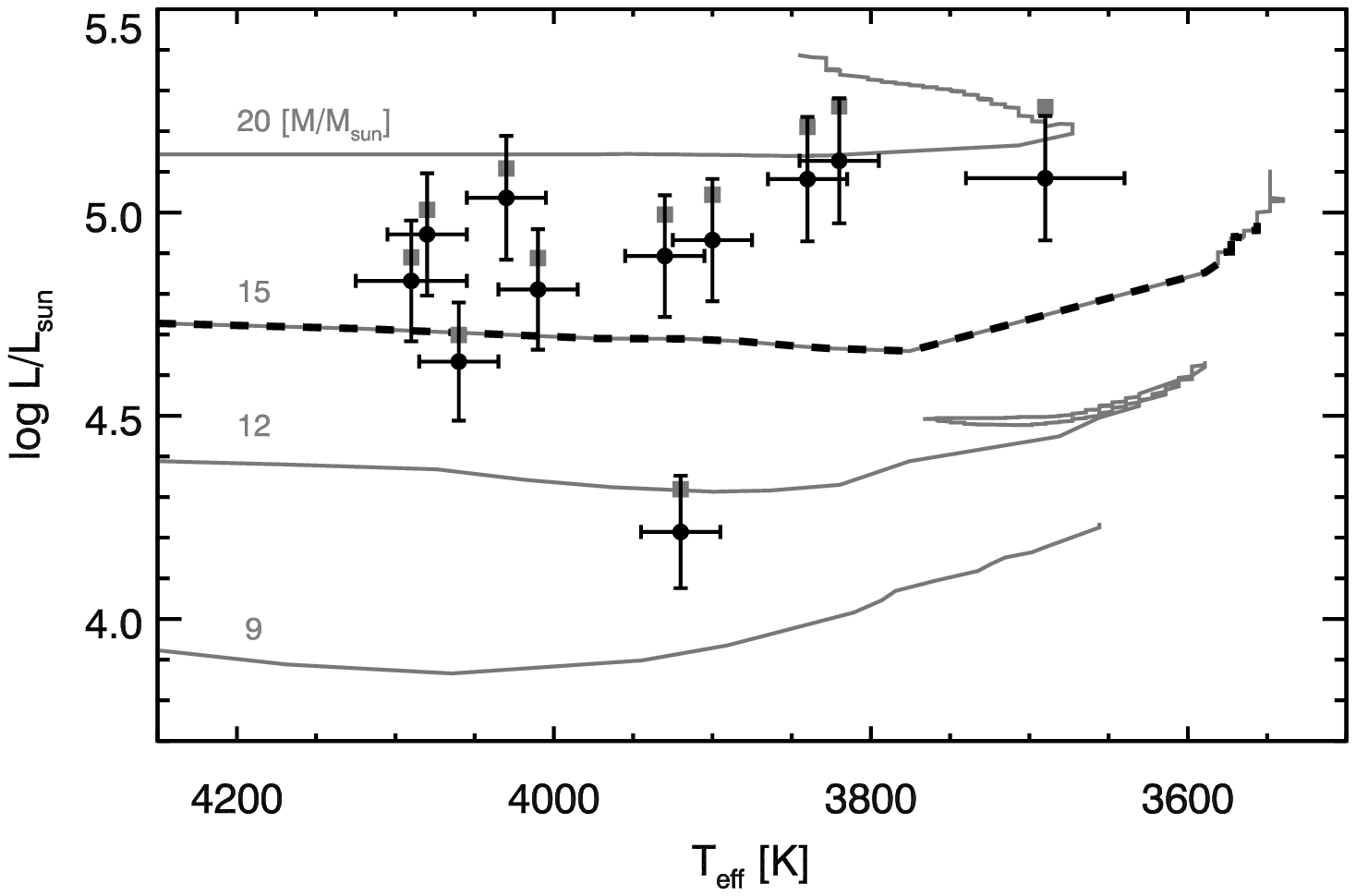}
\caption{H-R diagram of program stars.  Bolometric corrections are taken from \cite{2013ApJ...767....3D}.  Overplotted in gray are Geneva evolution tracks for solar metallicity including the effects of rotation, labeled with their zero-age main sequence mass.  The bold dashed overlay represents the space on the geneva tracks which covers the literature age of \obs, 14 $\pm$ 1 Myr \citep{2010ApJS..186..191C}. Gray squares show luminosities calculated using the bolometric corrections of \cite{2005ApJ...628..973L}, which are systematically higher but within 1$\sigma$.}
\label{fig:hrd}
\end{centering}
\end{figure}

To compare our results with stellar evolution models we 
first construct an observational Hertzsrung-Russel diagram
(HRD). We calculate bolometric luminosities for program stars 
using archival K band 2MASS photometry (Table~\ref{tbl:perOB1}, 
\citealt{2006AJ....131.1163S}), the bolometric correction recipes 
of \cite{2013ApJ...767....3D} and \cite{2005ApJ...628..973L}, and 
distance modulus of \cite{2010ApJS..186..191C}.  
We applied the \cite{Cardelli:1989p3360} extinction law using measurements 
of the reddening to \obs\ \citep{2010ApJS..186..191C}.  These luminosities are 
plotted against the effective temperatures from our 
spectral fit in the HRD of Figure~\ref{fig:hrd}. We then over plot 
evolutionary tracks adopting the Geneva database of 
stellar evolutionary models including the effects of 
rotation \citep{2000A&A...361..101M}. All stars except one align along the 
evolutionary tracks, having zero age main sequence 
(ZAMS) masses of 15-20 M$_{\odot}$. This result is in good
agreement with the age of \obs\ of 14$\pm$1 Myr \citep{2010ApJS..186..191C}. 
Only the 15 M$_{\odot}$ track agrees with
this time frame. At the age of \obs, 20 M$_{\odot}$ stars
have already evolved from the RSG phase while 
12 M$_{\odot}$ stars are still on the main sequence.

The object BD+59 372 is a clear 
outlier with a luminosity corresponding to a mass only
slightly higher than 9 M$_{\odot}$ (see Table~\ref{tbl:fitsnlte}). At this point we 
have no explanation for this object.  

\begin{figure}
\begin{centering}
\includegraphics[width=8.5cm]{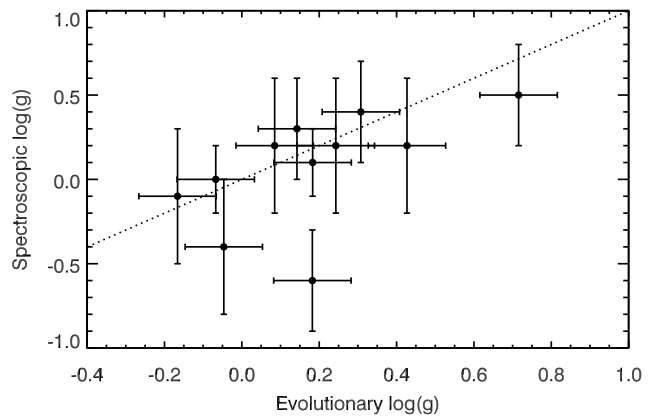}
\caption{A comparison between parameter fits for \logg\ and a calculation of the expected \logg\ values from stellar evolution theory given the age of \obs.  We find general agreement and note that outliers may be affected by significant turbulent pressure (see \S\ref{sec:stev})}
\label{fig:gvg}
\end{centering}
\end{figure}

An independent way to compare our spectroscopic results
with stellar evolution is the comparison of gravities
log g obtained from the spectroscopy and from evolutionary
tracks. For the latter, we obtain a stellar mass from 
the observed luminosities by interpolating evolutionary
track masses and luminosities at the effective temperature
observed. This mass is then used in conjunction with the
observed luminosity and effective temperature to calculate
evolutionary gravities. Figure~\ref{fig:gvg} compares evolutionary 
gravities obtained in this way with spectroscopic 
gravities. Besides one outlier (HD236979) we find general
agreement and no indication of a systematic discrepancy.
We also note that the outlier in Figure~\ref{fig:hrd}, BD+59 372, as the
object with the highest gravities agrees within the 
uncertainties of the error bars.

The general agreement between spectroscopic and 
evolutionary gravities can be used to discuss the 
influence of convective turbulence pressure on the model 
atmosphere stratification. The 3D-hydrodynamic convection 
simulations by \cite{2011A&A...535A..22C} include effects of
pressure caused by the convective motion on the 
atmospheric density stratification. On the other hand, the
1D MARCS models used in our analysis do not account for
convective pressure. It is straightforward to show (see,
for instance, \citealt{2011A&A...535A..22C}, eq. (8)) that as the
result of convective pressure the stellar gravity is 
reduced to an effective gravity which can be approximated
by

\begin{equation}
{\rm{log}}\,g_{\rm{eff}} = {\rm{log}}\,g - {\rm{log}}(1+\beta v_{\rm{turb}}^{2}/v_{\rm{sound}}^{2})
\end{equation}

where v$_{\rm{turb}}$ is the average turbulence speed and 
v$_{\rm{sound}}$ the sound speed. $\beta$ is a parameter close 
to unity if the turbulent velocity fields is almost 
isotropic. \cite{2011A&A...535A..22C} concluded from their
calculations and a comparison with MARCS models that 
gravity corrections of 0.25 to 0.3 dex are needed to match
the density stratifications of the 1D with the 3D models
corresponding to turbulence velocities of the order of the
sound speed.

Our comparison of spectroscopic and evolutionary gravities
does not indicate a systematic effect of this order. On 
the other hand, our two objects with the lowest 
spectroscopic gravities may well be influenced by large
effects of turbulence pressure.  We note, however, that 
the models used to calculate our synthetic spectra and
those models used for the stellar evolution calculations of
\citep{2000A&A...361..101M} utilize 1D models 
which may affect stellar evolution and atmosphere 
predictions in the same way.

\begin{figure*}
\begin{centering}
\includegraphics[width=8.5cm]{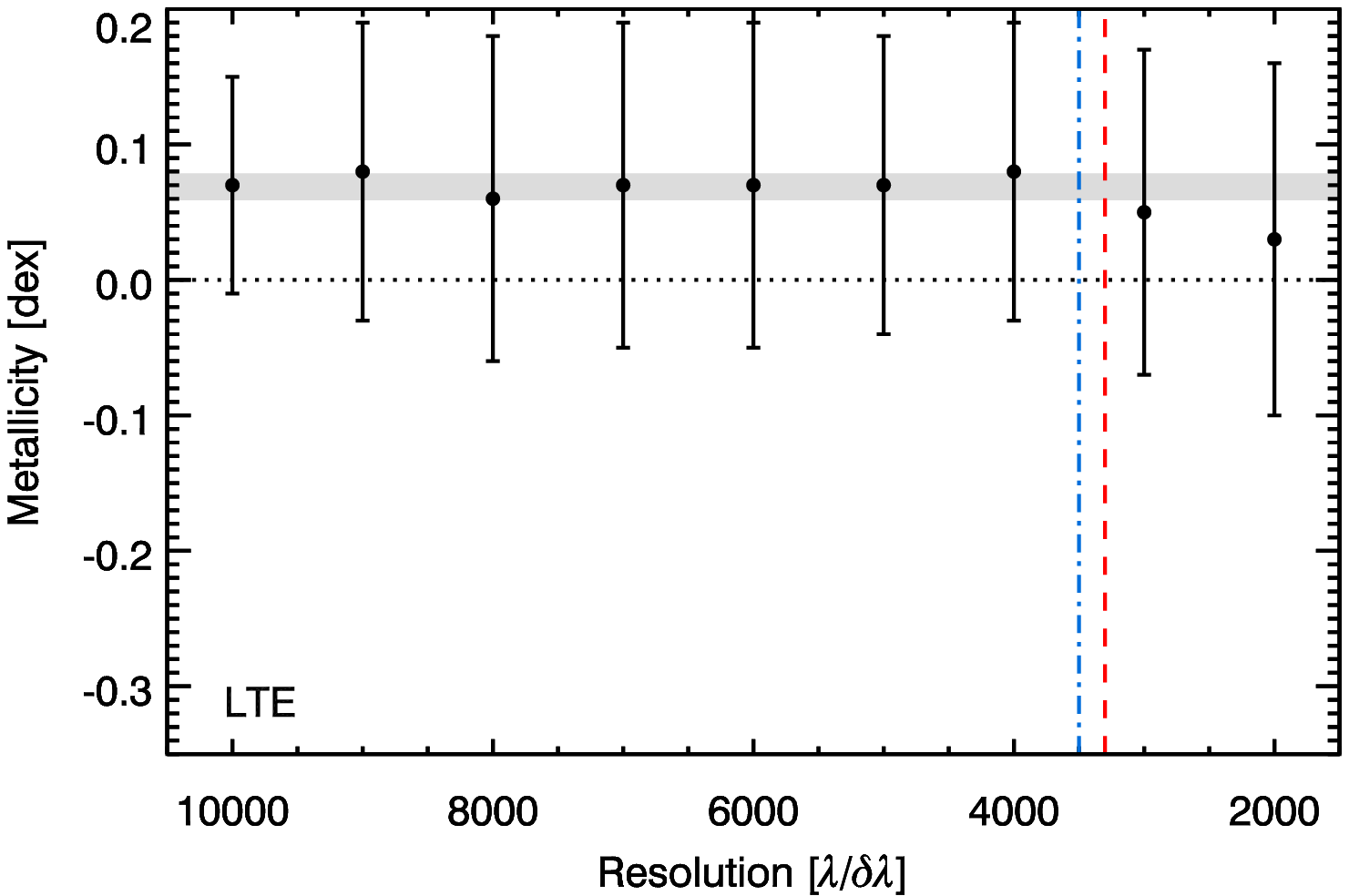} 
\includegraphics[width=8.5cm]{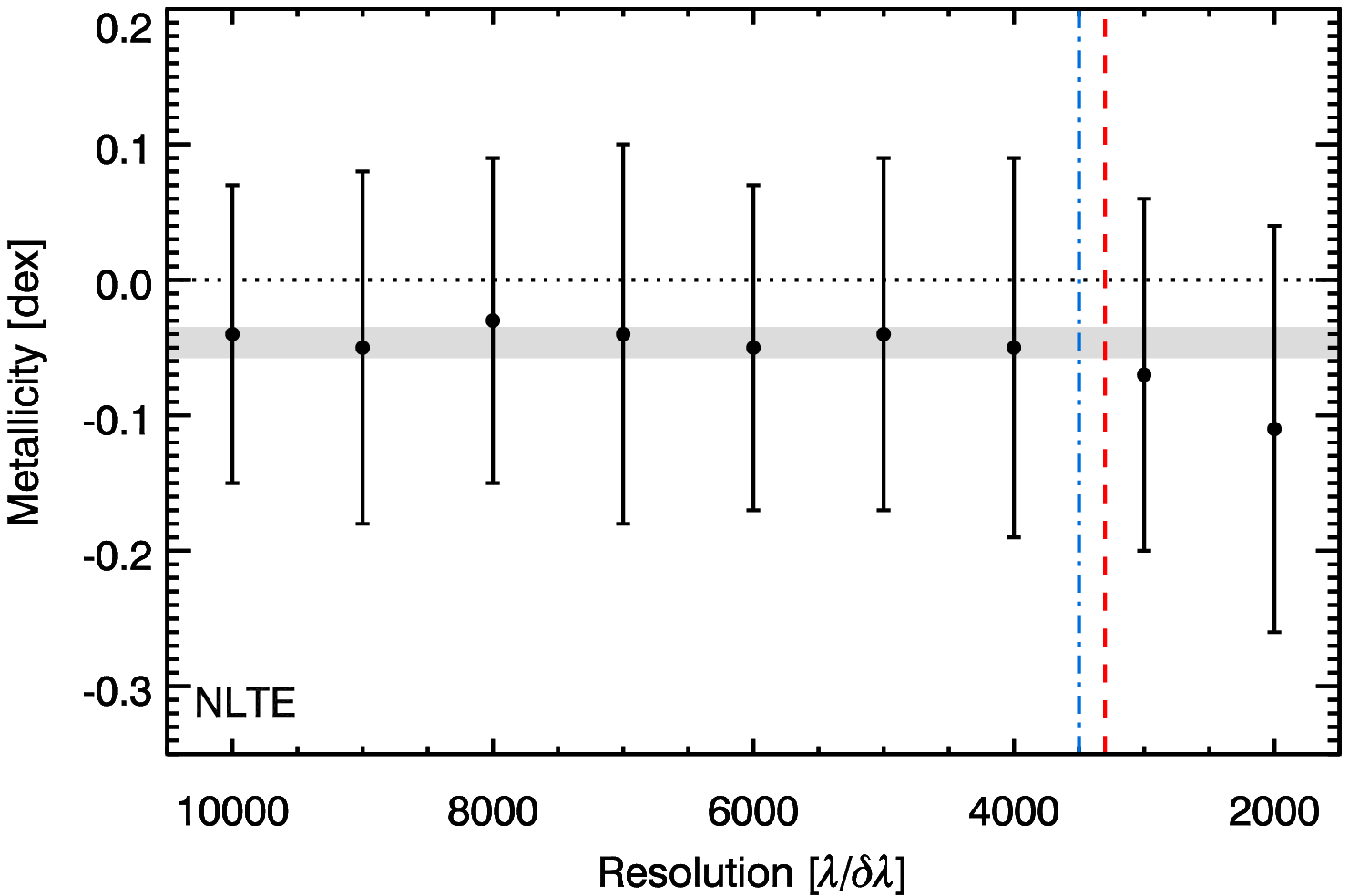} 
\caption{Evolution of the average measured metallicity for our sample of \obs\ stars collapsed into a synthetic cluster spectrum as a function of spectral resolution.   Error bars are derived using the Monte Carlo technique discussed in \S\ref{sec:parerr}.  The horizontal gray region shows $\pm$1$\sigma$ of the average metallicity between 10000 $\le$ R $\le $ 3000, demonstrating the stability of the technique down to resolutions of R=3000.  Vertical lines mark the spectral resolutions of key \jband\ spectrographs, KMOS on VLT in dash-dotted blue and MOSFIRE on Keck in dashed red.  A horizontal dotted line marks solar metallicity.  We plot results from the \lte\ model grid (left panel) and \nlte\ grid (right panel). }
\label{fig:zvrc}
\end{centering}
\end{figure*}

\subsection{Simulation of Super Star Cluster Spectral Analysis}

The scientific strength of the low resolution \jband\ technique derives 
from the radiative power of RSG stars.  In this work we carefully 
demonstrate that the method is stable and precise well below the 
spectral resolution of current instrumentation on the largest telescopes 
available, notably MOSFIRE on Keck and KMOS on the VLT.   With 
these multiplexed instruments we are able to efficiently apply this 
technique to entire populations of RSGs as individual objects over 
extragalactic distances.  DFK10 calculate a limiting distance for the 
technique of 7-10 MPC  using a single RSG.

In \cite{2013MNRAS.430L..35G} we presented simulations showing
that the near-IR flux of young super star clusters (SSCs)
is dominated by their RSG members. These simulations show
that the \jband\ spectrum of a SSC older than 7 Myr will 
appear very similar to that of a single RSGs. This opens
the possibility to use the integrated \jband\ light of SSCs
in distant galaxies as as a source for spectroscopic 
determination of galaxy metallicities.   

Our collection of \obs\ spectra allows us to test this
possibility. With a total estimated mass of 20000 
M$_{\odot}$ \citep{2010ApJS..186..191C} \obs\ comes close 
to the observed mass range of 
extragalactic SSCs. We construct a simulated SSC spectrum by
adding our observed RSGs weighted by their \jband\ 
luminosities. The spectrum is shown in Figure~\ref{fig:clusts}. We then apply
the same analysis technique as in section 3 and obtain a
metallicity of \met\ = \clustnlte\ (\nlte), very similar to
the average metallicity obtained from the analysis of the
11 individual spectra. The effective temperature obtained
from the cluster spectrum is \teff = 3970 $\pm$ 30 and the 
gravity \logg = $+$0.1 $\pm$ 0.2.  In agreement with the \lte\ study of the individual \obs\ supergiants we measured \met\ = \clustlte, \teff = 3910 $\pm$ 70, and \logg = +0.2  $\pm$ 0.1 when fitting with the full \lte\ model grid.  
 
\cite{2013MNRAS.430L..35G} find that the RSG supergiant population will provide $\sim$95\% of the \jband\ flux in a young SSC.  
To simulate the effect of the 5\% contaminative flux we added a flat spectrum of 5\% of the total flux.  We then re-fit the spectrum and measured \clustnltecon\ (\nlte) and \clustltecon\ (\lte).  The change in measured metallicity is minimal (with a systematic offset of at most $+$0.05 dex) and in the proper direction$-$contaminant flux will weaken the deepest lines more strongly and thus a drop in extracted metallicity is to be expected.  However, the two results agree statistically and offer strong evidence that spectroscopy of unresolved young SSCs can become a powerful application of the \jband\ technique.  We find in this case that an unresolved cluster of proper age can be successfully fit with a single RSG template model, a technique which has been used at very high resolution in the H band \citep{2006MNRAS.368L..10L}.  

We scale the resolution of our synthetic cluster spectrum down to R=2000.  The results of this work echo that of the individual stars, showing stability in fit parameters down to resolutions around 3000.  The NLTE and LTE cases of this test are plotted in Figure~\ref{fig:zvrc}.  

\section{conclusions}

In this paper we have tested the \jband\ technique for extracting metallicity information from modest resolution spectra of RSGs.  Through a careful suite of tests we have demonstrated the precision and accuracy of the technique.  We obtain reliable abundances in agreement with high resolution, high signal to noise spectroscopy of young massive B-stars in the solar neighborhood.  Using the advantage that all of our RSGs formed within a stellar cluster we test our derived parameters against predictions of stellar evolution theory for a cluster of mass and age of \obs.  Our results are in good agreement with such theoretical work.  We thus confirm the technique presented in DFK10 and show that it remains stable down to resolutions of R$\approx$3000.  This provides a reliable method to determine extragalactic metallicities from individual RSGs to distances of 7-10 Mpc with existing telescopes and instruments.  Both Keck (MOSFIRE) and the VLT (KMOS) have multi object spectrographs in the near-IR which operate above resolutions of 3000.  With these instruments and the \jband\ technique, RSGs across the entire disks of star forming galaxies can be observed efficiently.

By utilizing the large populations of RSGs in young, spatially unresolved SSCs we can extend the applicability of the \jband\ technique out to distances ten times greater with the same instruments.  Thus SSCs may allow us to reach beyond the local group and measure the metallicities of star forming galaxies from the stars themselves instead of relying on existing techniques which are empirically calibrated.  

We note that low resolution work is now needed in targets expected to be sub solar and super solar in metallicity.  The successful application of the \jband\ technique in such cases would pose the methods tested in this paper to study the metallicity evolution of star forming galaxies in a large volume of the nearby universe.

\section{Acknowledgments}
JZG and RPK acknowledge support by the National Science Foundation under grant AST-1108906 and the hospitality of the Munich University Observatory where part of this work was carried out.  This work was partly supported by the European Union FP7 programme through ERC grant number 320360.  BD is supported by a fellowship from the Royal Astronomical Society. BP is supported in part by the Programme National de Physique Stellaire of the INSU CNRS.  We thank our referee, F. Najarro, for improvements to this manuscript based on his careful review.


\end{document}